\documentclass[twocolumn,letterpaper,aps,prc,superscriptaddress,nofootinbib,showpacs,floatfix]{revtex4-1}

\usepackage{graphicx}
\usepackage{amssymb}
\usepackage{amsmath}
\usepackage{xspace}	
\usepackage{dcolumn}
\usepackage{bm}

\newcommand{\sqs}{\mbox{$\sqrt{s}$}\xspace}
\newcommand{\sqsn}{\mbox{$\sqrt{s_{_{NN}}}$}\xspace}
\newcommand{\energy}{\mbox{$\sqrt{s_{_{NN}}}=200$~GeV}\xspace}
\newcommand{\energypp}{\mbox{$\sqrt{s}=200$\,GeV}\xspace}
\newcommand{\pt}{\mbox{$p_T$}\xspace}
\newcommand{\pT}{\pt}
\newcommand{\gevc}{\mbox{GeV/$c$}\xspace}
\newcommand{\pp}{\mbox{$p$$+$$p$}\xspace}
\newcommand{\dau}{\mbox{$d$$+$Au}\xspace}
\newcommand{\ppb}{\mbox{$p$$+$Pb}\xspace}
\newcommand{\pbpb}{\mbox{Pb$+$Pb}\xspace}
\newcommand{\auau}{\mbox{Au$+$Au}\xspace}
\newcommand{\cucu}{\mbox{Cu$+$Cu}\xspace}
\newcommand{\raa}{\mbox{$R_{AA}$}\xspace}
\newcommand{\rab}{\mbox{$R_{AB}$}\xspace}
\newcommand{\rcucu}{\mbox{$R_{\rm CuCu}$}\xspace}

\newcommand{\rda}{\mbox{$R_{d\rm{Au}}$}\xspace}

\newcommand{\Np}{\mbox{$N_{\rm part}$}\xspace}
\newcommand{\Nc}{\mbox{$N_{\rm coll}$}\xspace}

\newcommand{\kstar} {\mbox{$K^{*0}$}\xspace}
\newcommand{\kstarbar} {\mbox{$\overline{K^{*0}}$}\xspace}
\newcommand{\kshort} {\mbox{$K_S^0$}\xspace}

\newcommand{\ksdecay} {\mbox{$K_S^0 \rightarrow \pi^0(\rightarrow \gamma\gamma)\pi^0(\rightarrow\gamma\gamma)$}\xspace}
\newcommand{\kstdecay} {\mbox{$K^{*0} \rightarrow K^{\pm}\pi^{\mp}$}\xspace}
\newcommand{\piz} {\mbox{$\pi^0$}\xspace}
\newcommand{\g} {\mbox{$\gamma$}\xspace}
\newcommand{\kpi} {\mbox{$K\pi$}\xspace}

\begin{document}

\title{Measurement of $K_S^0$ and $K^{*0}$ in $p$$+$$p$, $d$$+$Au, and 
Cu$+$Cu collisions at $\sqrt{s_{_{NN}}}=200$~GeV}

\newcommand{\abilene}{Abilene Christian University, Abilene, Texas 79699, USA}
\newcommand{\augie}{Department of Physics, Augustana College, Sioux Falls, South Dakota 57197, USA}
\newcommand{\banaras}{Department of Physics, Banaras Hindu University, Varanasi 221005, India}
\newcommand{\barc}{Bhabha Atomic Research Centre, Bombay 400 085, India}
\newcommand{\baruch}{Baruch College, City University of New York, New York, New York, 10010 USA}
\newcommand{\bnlcoll}{Collider-Accelerator Department, Brookhaven National Laboratory, Upton, New York 11973-5000, USA}
\newcommand{\bnlphys}{Physics Department, Brookhaven National Laboratory, Upton, New York 11973-5000, USA}
\newcommand{\caucr}{University of California - Riverside, Riverside, California 92521, USA}
\newcommand{\charlesczech}{Charles University, Ovocn\'{y} trh 5, Praha 1, 116 36, Prague, Czech Republic}
\newcommand{\chonbuk}{Chonbuk National University, Jeonju, 561-756, Korea}
\newcommand{\ciae}{Science and Technology on Nuclear Data Laboratory, China Institute of Atomic Energy, Beijing 102413, P.~R.~China}
\newcommand{\cns}{Center for Nuclear Study, Graduate School of Science, University of Tokyo, 7-3-1 Hongo, Bunkyo, Tokyo 113-0033, Japan}
\newcommand{\colorado}{University of Colorado, Boulder, Colorado 80309, USA}
\newcommand{\columbia}{Columbia University, New York, New York 10027 and Nevis Laboratories, Irvington, New York 10533, USA}
\newcommand{\czechtech}{Czech Technical University, Zikova 4, 166 36 Prague 6, Czech Republic}
\newcommand{\dapnia}{Dapnia, CEA Saclay, F-91191, Gif-sur-Yvette, France}
\newcommand{\debrecen}{Debrecen University, H-4010 Debrecen, Egyetem t{\'e}r 1, Hungary}
\newcommand{\elte}{ELTE, E{\"o}tv{\"o}s Lor{\'a}nd University, H - 1117 Budapest, P{\'a}zm{\'a}ny P. s. 1/A, Hungary}
\newcommand{\ewha}{Ewha Womans University, Seoul 120-750, Korea}
\newcommand{\fit}{Florida Institute of Technology, Melbourne, Florida 32901, USA}
\newcommand{\fsu}{Florida State University, Tallahassee, Florida 32306, USA}
\newcommand{\gsu}{Georgia State University, Atlanta, Georgia 30303, USA}
\newcommand{\hanyang}{Hanyang University, Seoul 133-792, Korea}
\newcommand{\hiroshima}{Hiroshima University, Kagamiyama, Higashi-Hiroshima 739-8526, Japan}
\newcommand{\howard}{Department of Physics and Astronomy, Howard University, Washington, DC 20059, USA}
\newcommand{\ihepprot}{IHEP Protvino, State Research Center of Russian Federation, Institute for High Energy Physics, Protvino, 142281, Russia}
\newcommand{\illuiuc}{University of Illinois at Urbana-Champaign, Urbana, Illinois 61801, USA}
\newcommand{\inrras}{Institute for Nuclear Research of the Russian Academy of Sciences, prospekt 60-letiya Oktyabrya 7a, Moscow 117312, Russia}
\newcommand{\instpasczech}{Institute of Physics, Academy of Sciences of the Czech Republic, Na Slovance 2, 182 21 Prague 8, Czech Republic}
\newcommand{\isu}{Iowa State University, Ames, Iowa 50011, USA}
\newcommand{\jaea}{Advanced Science Research Center, Japan Atomic Energy Agency, 2-4 Shirakata Shirane, Tokai-mura, Naka-gun, Ibaraki-ken 319-1195, Japan}
\newcommand{\jinrdubna}{Joint Institute for Nuclear Research, 141980 Dubna, Moscow Region, Russia}
\newcommand{\jyvaskyla}{Helsinki Institute of Physics and University of Jyv{\"a}skyl{\"a}, P.O.Box 35, FI-40014 Jyv{\"a}skyl{\"a}, Finland}
\newcommand{\kek}{KEK, High Energy Accelerator Research Organization, Tsukuba, Ibaraki 305-0801, Japan}
\newcommand{\korea}{Korea University, Seoul, 136-701, Korea}
\newcommand{\kurchatov}{Russian Research Center ``Kurchatov Institute", Moscow, 123098 Russia}
\newcommand{\kyoto}{Kyoto University, Kyoto 606-8502, Japan}
\newcommand{\labllr}{Laboratoire Leprince-Ringuet, Ecole Polytechnique, CNRS-IN2P3, Route de Saclay, F-91128, Palaiseau, France}
\newcommand{\lahorelums}{Physics Department, Lahore University of Management Sciences, Lahore, Pakistan}
\newcommand{\lawllnl}{Lawrence Livermore National Laboratory, Livermore, California 94550, USA}
\newcommand{\losalamos}{Los Alamos National Laboratory, Los Alamos, New Mexico 87545, USA}
\newcommand{\lpc}{LPC, Universit{\'e} Blaise Pascal, CNRS-IN2P3, Clermont-Fd, 63177 Aubiere Cedex, France}
\newcommand{\lund}{Department of Physics, Lund University, Box 118, SE-221 00 Lund, Sweden}
\newcommand{\maryland}{University of Maryland, College Park, Maryland 20742, USA}
\newcommand{\mass}{Department of Physics, University of Massachusetts, Amherst, Massachusetts 01003-9337, USA }
\newcommand{\michigan}{Department of Physics, University of Michigan, Ann Arbor, Michigan 48109-1040, USA}
\newcommand{\muenster}{Institut fur Kernphysik, University of Muenster, D-48149 Muenster, Germany}
\newcommand{\muhlenberg}{Muhlenberg College, Allentown, Pennsylvania 18104-5586, USA}
\newcommand{\myongji}{Myongji University, Yongin, Kyonggido 449-728, Korea}
\newcommand{\nagasaki}{Nagasaki Institute of Applied Science, Nagasaki-shi, Nagasaki 851-0193, Japan}
\newcommand{\newmex}{University of New Mexico, Albuquerque, New Mexico 87131, USA }
\newcommand{\nmsu}{New Mexico State University, Las Cruces, New Mexico 88003, USA}
\newcommand{\ohio}{Department of Physics and Astronomy, Ohio University, Athens, Ohio 45701, USA}
\newcommand{\ornl}{Oak Ridge National Laboratory, Oak Ridge, Tennessee 37831, USA}
\newcommand{\orsay}{IPN-Orsay, Universite Paris Sud, CNRS-IN2P3, BP1, F-91406, Orsay, France}
\newcommand{\peking}{Peking University, Beijing 100871, P.~R.~China}
\newcommand{\pnpi}{PNPI, Petersburg Nuclear Physics Institute, Gatchina, Leningrad region, 188300, Russia}
\newcommand{\riken}{RIKEN Nishina Center for Accelerator-Based Science, Wako, Saitama 351-0198, Japan}
\newcommand{\rikjrbrc}{RIKEN BNL Research Center, Brookhaven National Laboratory, Upton, New York 11973-5000, USA}
\newcommand{\rikkyo}{Physics Department, Rikkyo University, 3-34-1 Nishi-Ikebukuro, Toshima, Tokyo 171-8501, Japan}
\newcommand{\saispbstu}{Saint Petersburg State Polytechnic University, St. Petersburg, 195251 Russia}
\newcommand{\saopaulo}{Universidade de S{\~a}o Paulo, Instituto de F\'{\i}sica, Caixa Postal 66318, S{\~a}o Paulo CEP05315-970, Brazil}
\newcommand{\seoulnat}{Department of Physics and Astronomy, Seoul National University, Seoul, Korea}
\newcommand{\stonybrkc}{Chemistry Department, Stony Brook University, SUNY, Stony Brook, New York 11794-3400, USA}
\newcommand{\stonycrkp}{Department of Physics and Astronomy, Stony Brook University, SUNY, Stony Brook, New York 11794-3800, USA}
\newcommand{\subatech}{SUBATECH (Ecole des Mines de Nantes, CNRS-IN2P3, Universit{\'e} de Nantes) BP 20722 - 44307, Nantes, France}
\newcommand{\tenn}{University of Tennessee, Knoxville, Tennessee 37996, USA}
\newcommand{\titech}{Department of Physics, Tokyo Institute of Technology, Oh-okayama, Meguro, Tokyo 152-8551, Japan}
\newcommand{\tsukuba}{Institute of Physics, University of Tsukuba, Tsukuba, Ibaraki 305, Japan}
\newcommand{\vandy}{Vanderbilt University, Nashville, Tennessee 37235, USA}
\newcommand{\waseda}{Waseda University, Advanced Research Institute for Science and Engineering, 17 Kikui-cho, Shinjuku-ku, Tokyo 162-0044, Japan}
\newcommand{\weizmann}{Weizmann Institute, Rehovot 76100, Israel}
\newcommand{\wigner}{Institute for Particle and Nuclear Physics, Wigner Research Centre for Physics, Hungarian Academy of Sciences (Wigner RCP, RMKI) H-1525 Budapest 114, POBox 49, Budapest, Hungary}
\newcommand{\yonsei}{Yonsei University, IPAP, Seoul 120-749, Korea}
\newcommand{\zagreb}{University of Zagreb, Faculty of Science, Department of Physics, Bijeni\v{c}ka 32, HR-10002 Zagreb, Croatia}
\affiliation{\abilene}
\affiliation{\augie}
\affiliation{\banaras}
\affiliation{\barc}
\affiliation{\baruch}
\affiliation{\bnlcoll}
\affiliation{\bnlphys}
\affiliation{\caucr}
\affiliation{\charlesczech}
\affiliation{\chonbuk}
\affiliation{\ciae}
\affiliation{\cns}
\affiliation{\colorado}
\affiliation{\columbia}
\affiliation{\czechtech}
\affiliation{\dapnia}
\affiliation{\debrecen}
\affiliation{\elte}
\affiliation{\ewha}
\affiliation{\fit}
\affiliation{\fsu}
\affiliation{\gsu}
\affiliation{\hanyang}
\affiliation{\hiroshima}
\affiliation{\howard}
\affiliation{\ihepprot}
\affiliation{\illuiuc}
\affiliation{\inrras}
\affiliation{\instpasczech}
\affiliation{\isu}
\affiliation{\jaea}
\affiliation{\jinrdubna}
\affiliation{\jyvaskyla}
\affiliation{\kek}
\affiliation{\korea}
\affiliation{\kurchatov}
\affiliation{\kyoto}
\affiliation{\labllr}
\affiliation{\lahorelums}
\affiliation{\lawllnl}
\affiliation{\losalamos}
\affiliation{\lpc}
\affiliation{\lund}
\affiliation{\maryland}
\affiliation{\mass}
\affiliation{\michigan}
\affiliation{\muenster}
\affiliation{\muhlenberg}
\affiliation{\myongji}
\affiliation{\nagasaki}
\affiliation{\newmex}
\affiliation{\nmsu}
\affiliation{\ohio}
\affiliation{\ornl}
\affiliation{\orsay}
\affiliation{\peking}
\affiliation{\pnpi}
\affiliation{\riken}
\affiliation{\rikjrbrc}
\affiliation{\rikkyo}
\affiliation{\saispbstu}
\affiliation{\saopaulo}
\affiliation{\seoulnat}
\affiliation{\stonybrkc}
\affiliation{\stonycrkp}
\affiliation{\subatech}
\affiliation{\tenn}
\affiliation{\titech}
\affiliation{\tsukuba}
\affiliation{\vandy}
\affiliation{\waseda}
\affiliation{\weizmann}
\affiliation{\wigner}
\affiliation{\yonsei}
\affiliation{\zagreb}
\author{A.~Adare} \affiliation{\colorado}
\author{S.~Afanasiev} \affiliation{\jinrdubna}
\author{C.~Aidala} \affiliation{\columbia} \affiliation{\mass} \affiliation{\michigan}
\author{N.N.~Ajitanand} \affiliation{\stonybrkc}
\author{Y.~Akiba} \affiliation{\riken} \affiliation{\rikjrbrc}
\author{R.~Akimoto} \affiliation{\cns}
\author{H.~Al-Bataineh} \affiliation{\nmsu}
\author{J.~Alexander} \affiliation{\stonybrkc}
\author{M.~Alfred} \affiliation{\howard}
\author{A.~Angerami} \affiliation{\columbia}
% ...
\collaboration{PHENIX Collaboration} \noaffiliation

\date{\today}  


\begin{abstract}


The PHENIX experiment at the Relativistic Heavy Ion Collider has performed 
a systematic study of $K_S^0$ and $K^{*0}$ meson production at midrapidity 
in $p$$+$$p$, $d$$+$Au, and Cu$+$Cu collisions at 
$\sqrt{s_{_{NN}}}=200$~GeV. The $K_S^0$ and $K^{*0}$ mesons are 
reconstructed via their
$K_S^0 \rightarrow \pi^0(\rightarrow \gamma\gamma)\pi^0(\rightarrow\gamma\gamma)$
and $K^{*0} \rightarrow K^{\pm}\pi^{\mp}$ decay modes, respectively.  The 
measured transverse-momentum spectra are used to determine the nuclear 
modification factor of $K_S^0$ and $K^{*0}$ mesons in $d$$+$Au and Cu$+$Cu 
collisions at different centralities. In the $d$$+$Au collisions, the 
nuclear modification factor of $K_S^0$ and $K^{*0}$ mesons is almost 
constant as a function of transverse momentum and is consistent with unity 
showing that cold-nuclear-matter effects do not play a significant role in 
the measured kinematic range. In Cu$+$Cu collisions, within the 
uncertainties no nuclear modification is registered in peripheral 
collisions.  In central collisions, both mesons show suppression relative 
to the expectations from the $p$$+$$p$ yield scaled by the number of 
binary nucleon-nucleon collisions in the Cu$+$Cu system. In the $p_T$ 
range 2--5~GeV/$c$, the strange mesons ($K_S^0$, $K^{*0}$) similarly to 
the $\phi$ meson with hidden strangeness, show an intermediate suppression 
between the more suppressed light quark mesons ($\pi^0$) and the 
nonsuppressed baryons ($p$, $\bar{p}$). At higher transverse momentum, 
$p_T>5$~GeV/$c$, production of all particles is similarly suppressed by a 
factor of $\approx$\,2.

\end{abstract}


\maketitle

\section{Introduction}
\label{sec:Intro}

At very high energy densities, exceeding approximately 1\,GeV/fm$^3$, 
quantum chromodynamics predicts a phase transition from ordinary 
hadronic nuclear matter to a new state of matter where the degrees of 
freedom are quarks and gluons~\cite{QGP_shuryak}. This state of matter 
exhibits very strong coupling between its constituents and is thus called 
the strongly coupled Quark-Gluon Plasma (sQGP)~\cite{QGP}. Matter at such 
high energy density can be produced in laboratory conditions by colliding 
heavy nuclei at relativistic energies. A wealth of measurements is 
available from the experiments at the Relativistic Heavy Ion Collider 
(RHIC) and recently from the experiments at the Large Hadron Collider 
(LHC)~\cite{review}.

High-momentum penetrating probes are among the observables attracting 
primary attention. Highly energetic partons traversing the sQGP medium 
suffer significant energy loss~\cite{ELOSS1, ELOSS2}, leading to 
modification of the fragmentation functions~\cite{theory_ff} and softening 
of the measured transverse momentum (\pT) distribution. The softening of 
the spectrum is quantified by the ``nuclear modification factor'' (\rab) 
defined as:
\begin{eqnarray}
  \rab =  \frac{d^2 N_{AB}/dydp_{T}}{\Nc \times d^2 N_{pp}/dydp_{T}},
\label{eq:rAA}
\end{eqnarray}
where the numerator is the per-event yield of particle production in 
$A$+$B$ (heavy ion) collisions, measured as a function of \pt, $d^2 
N_{pp}/dydp_{T}$ is the per-event yield of the same process in \pp 
collisions and \Nc is the number of nucleon-nucleon collisions in the 
$A$+$B$ system~\cite{jetPHEN,PPG084pi0cucu}. \rab different from unity is 
a manifestation of medium effects. However, to untangle final state 
effects, such as energy loss, from possible contributions of cold nuclear 
matter and initial state effects (e.g. shadowing~\cite{SHADOW} and the 
Cronin effect ~\cite{CRONIN}), the nuclear modification factor must also 
be measured in systems like $p$$+A$ or $d$$+$$A$.

A significant suppression of hadrons produced in heavy ion collisions was 
first measured at RHIC~\cite{Phenix_hadrons_130, Phenix_hadrons_cent_130, 
Phenix_pi0_200, Phenix_chargedhadrons_200, Phenix_eta_auau, phipaper, 
Star_cent_auau_130, Star_hadron_corr_200, Star_hadron_supp_auau, 
Star_ch_part} and recently at the LHC~\cite{cms_chargedparticle_pbpb_2760, 
alice_chargedparticle_pbpb_2760} also with fully reconstructed 
jets~\cite{CMS_jets, CMS_dijets, ATLAS_jets}. In central \auau collisions 
at RHIC, \rab of hadrons reaches a maximum suppression of a factor of 
$\sim$ 5 at $\pt\sim5$\,\gevc~\cite{Phenix_pi0_200, Phenix_eta_auau, 
phipaper, omegapaper}. At higher \pt, the suppression is found to be 
independent of the particle type, mesons or baryons, and their quark 
flavor content~\cite{SINGE1, SINGE2, STAR_highpT}. In central \pbpb 
collisions at the LHC, the suppression reaches a factor of $\sim$ 7 at \pt 
$\sim$\,6--7~\gevc~\cite{cms_chargedparticle_pbpb_2760,alice_chargedparticle_pbpb_2760}. 
At higher \pt, the \rab starts to increase reaching a value of 0.5 at 
$p_T>40$\,\gevc.

In the intermediate \pt range ($2<p_T<5$\,\gevc), mesons containing 
light quarks ($\pi$, $\eta$) exhibit 
suppression~\cite{PPG146, Phenix_eta_auau}, whereas protons show very 
little or no suppression~\cite{PPG146, PPG015, PPG026}. Other processes, 
such as the Cronin effect~\cite{CRONIN}, strong radial 
flow~\cite{RADFLOW}, recombination effects~\cite{reco} have been invoked 
to explain the differences between mesons and baryons in this momentum 
range. Recent results obtained at the LHC in \ppb 
collisions~\cite{lhc_flow_cms, Atlas_conf, Alice_charged_p_pb} and at RHIC 
in \dau collisions~\cite{PPG146, PPG152} suggest that collective effects 
might be present even in small systems and can significantly modify the 
particle properties in the intermediate transverse momentum range.

Measurements of particles with different quark content provide additional 
constraints on the models of collective behavior, parton energy loss and 
parton recombination.  Experimental measurements of particles containing 
strange quarks are important to find out whether flow or recombination 
mechanisms boost strange hadron production at intermediate \pt and to 
understand their suppression at high \pt. In heavy ion collisions, the 
$\phi$ meson~\cite{phipaper} shows at high \pt the same suppression as 
particles containing only $u$ and $d$ quarks, however at intermediate \pt 
it is less suppressed than the $\pi$ meson.  On the other hand, the $\eta$ 
meson, which has a significant strange quark content, is suppressed at the 
same level as $\pi$ meson in the \pt range from 
2--10~\gevc~\cite{Phenix_eta_auau}.  Open questions are:  Which physics 
mechanism prevails in the intermediate \pt region and which process is 
responsible for the suppression of particles with strange quark content.

This article presents results of the \kshort and \kstar meson production 
as a function of \pt at midrapidity in \pp, \dau and \cucu collisions at 
\sqsn = 200\,GeV. The present measurements significantly extend the \pt 
reach of the previous PHENIX results on the measurement of \kshort meson 
in \pp collisions~\cite{PPG099}. The \kshort meson is reconstructed via 
the \ksdecay decay mode. The \kstar and \kstarbar mesons are reconstructed 
via the $\kstar \rightarrow K^{+}\pi^{-}$ and $\kstarbar \rightarrow 
K^{-}\pi^{+}$ decay modes, respectively. The yields measured for the 
\kstar and \kstarbar mesons are averaged together and denoted as \kstar. 
The invariant transverse momentum spectra for \kshort mesons are measured 
over the \pt range of 2--13~(3--12)~\gevc in the \dau (\cucu) 
collision systems. The \kstar meson spectra are measured in the \pt range 
from 1.1~\gevc up to 8--8.5~\gevc depending on the collision system.  The 
measurements extend the momentum coverage of the previously published 
results by the STAR collaboration~\cite{starKstarpp, starKstardau, 
starKstarcucu}. The nuclear modification factors are obtained for both 
particles in \dau and \cucu collisions at different centralities and are 
compared with those of the $\phi$ and \piz mesons. The measured \pt ranges 
and the centrality bins used in the different systems are listed in 
Table~\ref{centrality_bins}.

\begin{table}[tbh]
  \caption[] {Summary of centrality bins and measured \pt ranges for the \kshort and \kstar studies.}
  \label{centrality_bins}
    \begin{ruledtabular}\begin{tabular}{cccc}
        &           & Centrality  & Measured \pt \\
        & Collision &  bins & range \\
        & System &  (\%) & (GeV/$c$)\\
\hline
\kshort & \dau  & 0--20, 20--40, 40--60, 60--88 & 2.0--13.0\\
        & \cucu & 0--20, 20--60, 60--94             & 3.0--12.0\\
\kstar  & \pp   &          --------------                     & 1.1--8.0\\
        & \dau  & 0--20, 20--40, 40--60, 60--88 & 1.1--8.5\\
        & \cucu & 0--20, 20--40,  40--60, 60--94 & 1.4--8.0\\
    \end{tabular}\end{ruledtabular}
\end{table}

The paper is organized as follows. The next section gives a brief 
description of the PHENIX detector. The analysis procedures used to 
measure \kshort and \kstar mesons are described in 
Section~\ref{sec:Analysis}. The results, including the invariant \pt 
distributions and \rab, are given in Section~\ref{sec:Results}. A summary 
is given in Section~\ref{sec:Summary}.

\section{PHENIX Detector}
\label{sec:PHENIX Detector}

A detailed description of the PHENIX detector can be found in 
Ref.~\cite{PHENIX_overview}. The analysis here is performed using the two 
central-arm spectrometers, each covering an azimuthal angle $\phi=\pi/2$ 
and pseudorapidity $|\eta|<0.35$~\cite{PHENIX_tracking2} at midrapidity.  
Each arm comprises a Drift Chamber (DC), two or three layers of pad 
chambers (PC), a ring-imaging \v{C}erenkov detector (RICH) and an 
Electromagnetic Calorimeter (EMCal) and a time-of-flight detector (TOF). 
This analysis uses the east arm of the TOF detector that covers $\pi/4$ in 
$\phi$.

The global event information is provided by the beam-beam counters 
(BBC)~\cite{PHENIX_inner}, which are used for event triggering, collision 
time determination, measurement of the vertex position along the beam axis 
and for the centrality determination~\cite{centralitydAu, PPG084pi0cucu}. 
The typical vertex position resolution by the BBC depends on the track 
multiplicity and varies from $\sim$ 1.1\,cm in \pp collisions to $\sim$ 
3\,mm in central \auau collisions.

Track reconstruction in PHENIX is provided by two detectors: DC and 
PC~\cite{PHENIX_tracking2}. The DC and the first layer of PC (PC1) form 
the inner tracking system, whereas PC2 and PC3 form the outer tracker. The 
DC is a multiwire gaseous detector located outside the magnetic field 
between the radii of 2.02\,m and 2.48\,m in each PHENIX arm. The DC 
measures the track position with an angular resolution of $\sim$ 0.8\,mrad 
in the bending plane perpendicular to the beam axis. A combinatorial Hough 
Transform technique~\cite{hough_tr} is used to determine the track 
direction in azimuth and its bending angle in the axial magnetic field of 
the central magnet~\cite{PHENIX_magnet}. The track-reconstruction 
algorithm approximates all tracks in the volume of the DC with straight 
lines and assumes their origin at the collision vertex. This information 
is then combined with the hit information in PC1 which immediately follows 
the DC along the particle tracks. PC1 provides the $z$-coordinate 
information with a spatial resolution of $\sigma_z\sim$ 1.7\,mm. The 
resulting momentum resolution for charged particles with 
$p_T>0.2$\,\gevc is $\delta p / p$ = 0.7 $\oplus$ 1.1 \% $p$ (GeV/$c$), 
where the first term represents multiple scattering and the second term is due 
to the intrinsic angular resolution of the DC. Matching the tracks to hits 
in PC2 and PC3 located at radii of 4.2\,m and 5.0\,m respectively helps to 
reject secondary tracks that originate either from decays of long-lived 
hadrons or from interactions with the detector material. Detailed 
information on the PHENIX tracking can be found in 
Ref.~\cite{PHENIX_tracking, PHENIX_tracking2}.

The TOF detector~\cite{PHENIX_particle_id} identifies charged hadrons; 
pions, kaons and protons. It is located at a radial distance of 5.06\,m 
from the interaction point in the east central arm. The total timing 
resolution of TOF east is 130\,ps, which includes the start time 
determination from the BBC. This allows for a $2.6\sigma$ $\pi/K$ 
separation up to $\pt \simeq 2.5$\,\gevc and $K/p$ separation up to \pt = 
4.5\,\gevc using an asymmetric particle-identification (PID) cut, as 
described in Ref.~\cite{TOF_ppg101}.

The EMCal~\cite{PHENIX_calorimeter} uses lead-scintillator (PbSc) and 
lead-glass (PbGl) technologies and measures the position and energy of 
electrons and photons. It also provides a trigger on rare events with high 
momentum photons. The EMCal covers the full acceptance of the central 
spectrometers and is divided into eight sectors in azimuth. Six PbSc 
sectors are located at a radial distance of 5.1\,m from the beam line and 
comprise 15,552 lead-scintillator sandwich towers with cross section of 
$5.5\times5.5$\,cm$^{2}$ and depth of 18 radiation lengths ($X_{0}$). Two 
PbGl sectors are located at a distance of 5\,m and comprise 9,216 towers 
of $4\times4$\,cm$^{2}$ and a depth of 14.3\,$X_{0}$. Most electromagnetic 
showers extend over several towers. Groups of adjacent towers with signals 
above a threshold that are associated with the same shower form an EMCal 
cluster. The energy resolution of the PbSc (PbGl) calorimeter is $\delta 
E/E$ = 2.1 (0.8)\% $\oplus$ 8.1 (5.9)/$\sqrt{E[{\rm GeV}]}\%$. The spatial 
resolution of the PbSc (PbGl) calorimeter reaches $\sigma(E)$ = 1.55 (0.2) 
$\oplus$ 5.74 (8.4)/$\sqrt{E[{\rm GeV}]}$\,mm for particles at normal 
incidence.

Analyses presented in this paper use both the minimum bias (MB) and the 
rare event, EMCal-RICH trigger (ERT).  For \pp, \dau, and \cucu 
collisions, the MB trigger requires a coincidence of at least one channel 
firing on each side of the BBC. It further requires the vertex position 
along the beam axis $z$, as determined from the BBC timing information, to 
be within 38 cm of the nominal center of the interaction region. Photon 
ERT utilizes the EMCal to select events with at least one registered high 
\pt photon or electron. For every EMCal super 
module~\cite{PHENIX_calorimeter}, the ERT sums the registered 
energy in adjacent $4\times4$ EMCal towers. This trigger is used to 
collect samples for the \kshort meson analysis. The trigger fires if the 
summed energy exceeds 1.4 and 2.8 GeV threshold in \dau and \cucu 
collisions, respectively. The calculation of the ERT efficiency 
for photons and \kshort mesons is described in Section~\ref{inv_yields}.

\begin{figure*}[htb]
  \includegraphics[width=0.45\linewidth]{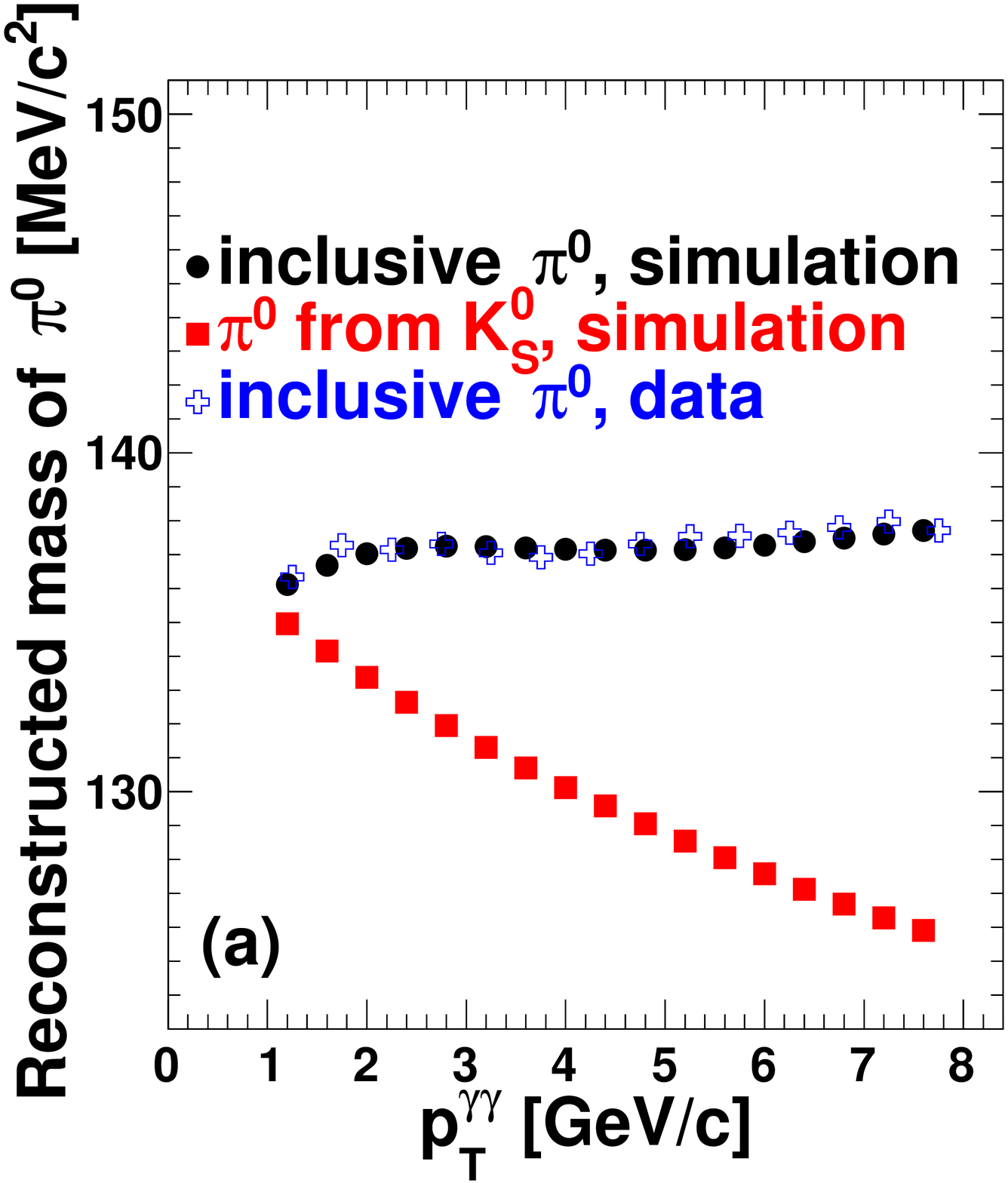}  
  \includegraphics[width=0.45\linewidth]{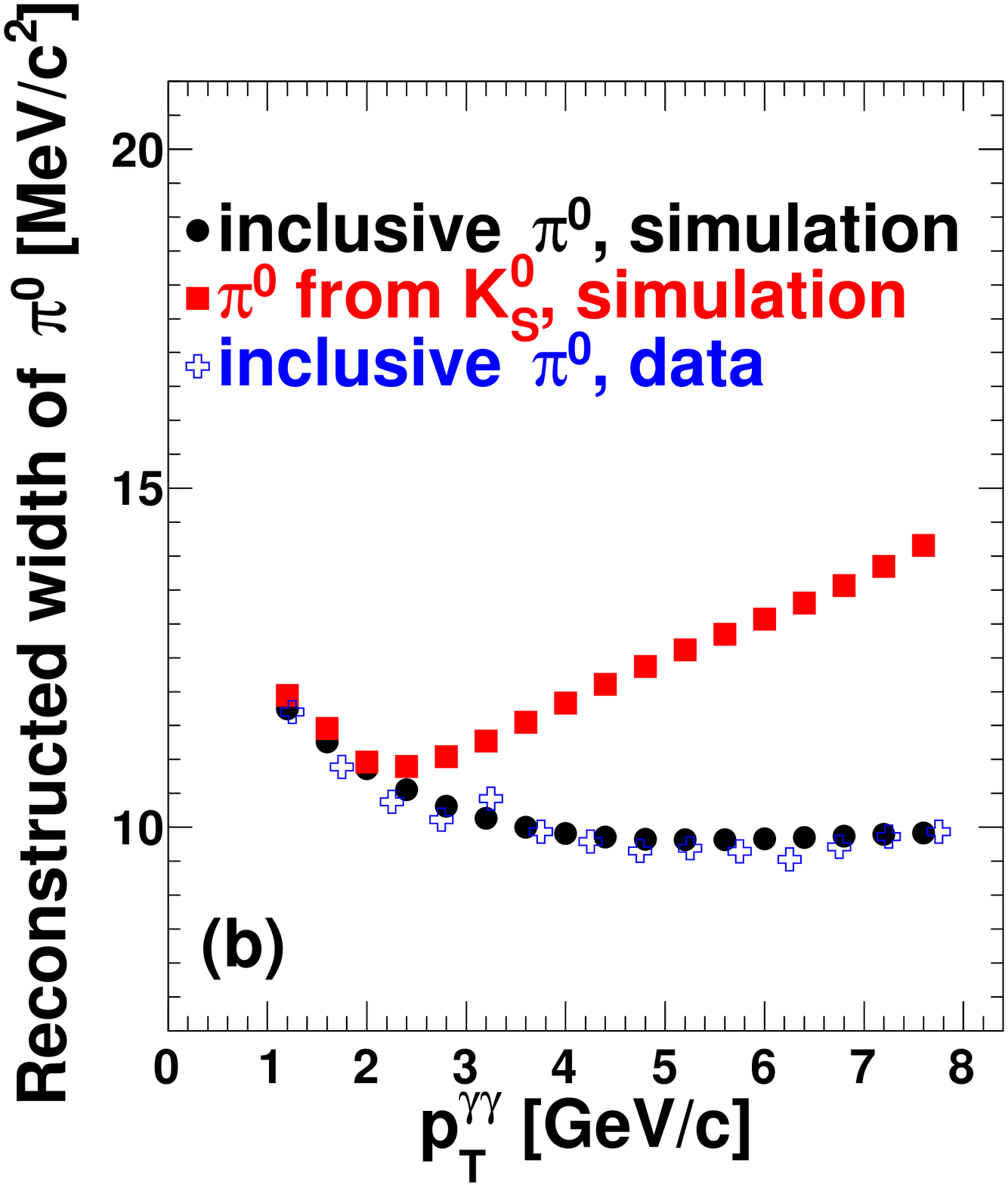} 
  \caption{(color online) 
(a) Reconstructed mass and (b) 1-$\sigma$ width of \piz as a function of 
the reconstructed \pt for inclusive \piz mesons from data (open crosses), 
simulations (circles) and for \piz coming from \kshort decays (squares).}
\label{fig:ksmasswidth}

  \begin{minipage}{0.5\linewidth}
\includegraphics[width=0.81\linewidth]{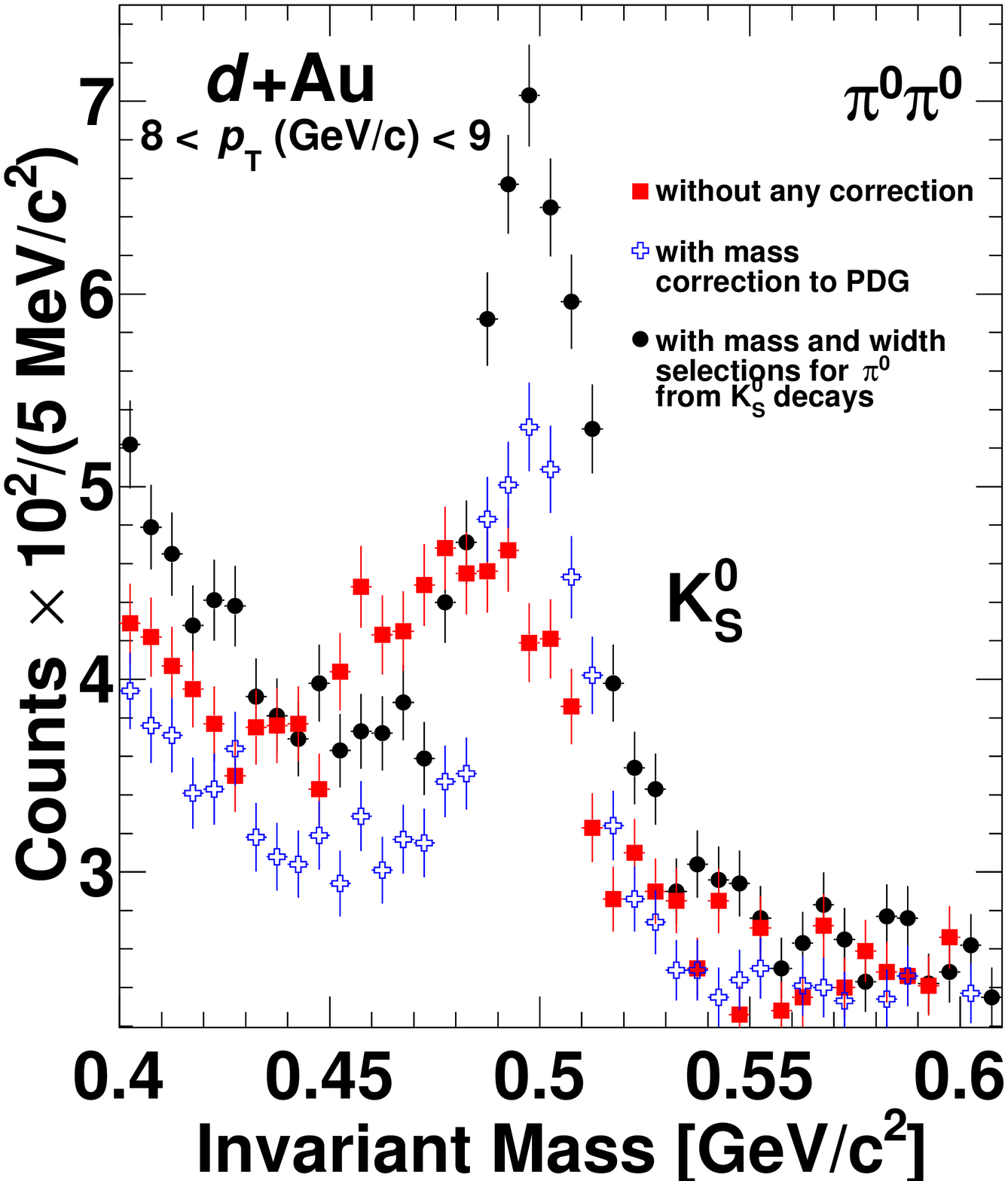}  
  \end{minipage}
  \hspace{0.05\linewidth}
  \begin{minipage}{0.3\linewidth}
\caption{(color online)
Invariant mass distribution for $\piz\piz$ pairs measured in the MB \dau 
collisions at $8<p_T<9$\,\gevc.  Invariant mass reconstructed without any 
corrections is shown with red squares. Invariant mass reconstructed after 
corrections for the mass of reconstructed \piz to the PDG value is shown 
with blue open crosses. Same with additional correction accounting for the 
difference between inclusive \piz mesons and neutral pions produced in 
\kshort meson decay as described in the text is shown with black circles.
}
\label{fig:ksinvmass_comp}
  \end{minipage}
\end{figure*}

\section{Analysis Procedure}
\label{sec:Analysis}

This section describes the analysis procedure for the measurement of 
\kshort meson and \kstar meson transverse momentum spectra. The 
measurements are done using the data sets collected by the PHENIX 
experiment in the 2005 (\pp and \cucu) and in the 2008 ($d$$+$Au) physics 
runs. The data samples used in the analysis correspond to integrated 
luminosities of 3.78 pb$^{-1}$ in \pp, 81 nb$^{-1}$ in \dau and 3.06 
nb$^{-1}$ in \cucu collision systems. The mesons are reconstructed via the 
decay modes \ksdecay and \kstdecay. The MB triggered data samples are used 
for the \kstar meson study in \pp, \dau and \cucu systems. The \kshort 
meson measurements are done using both the MB and ERT-triggered data 
samples in \dau and \cucu collisions. The MB samples provide the 
measurements at low and intermediate \pt. The low \pt reach of these 
measurements is limited by the rapidly decreasing signal to background 
ratio and subsequent difficulties in the extraction of the \kshort meson 
raw yield. The ERT-triggered data give access to intermediate and high \pt 
production of \kshort mesons due to larger sampled luminosity. In the 
overlap region, results obtained with the MB and ERT data samples are 
found to be in very good agreement. For the final \kshort meson production 
spectrum in \dau (\cucu) collisions, the MB results are used up to 4 (5) 
GeV/$c$ and the ERT results are used at higher transverse momenta.  
Details about the \kshort meson measurement in \pp collisions can be found 
in Ref.~\cite{PPG099}.

\subsection{Reconstruction of \kshort meson invariant mass}
\label{kshort_ana}

The \kshort meson with a lifetime of c$\tau\sim$ 2.7\,cm decays to two \piz 
mesons with a branching ratio BR = $30.69\pm0.05$\%~\cite{PDG}. The 
neutral pions further decay into two photons with BR = 
$98.823\pm0.034$\%~\cite{PDG}. The \piz mesons are measured by combining 
the pair of photon clusters reconstructed in the EMCal. The energy of the 
clusters is measured in the EMCal and momentum components are calculated 
assuming that the particle originates at the event vertex. Besides 
electromagnetic showers created by photons and electrons, the EMCal also 
registers showers associated with hadrons. Because hadron showers are 
typically wider than the electromagnetic ones, a shower profile 
cut~\cite{em_profile_cut} is used to reject hadron-like clusters. The 
shower profile cut is based on a comparison of the registered cluster 
energy distribution in the EMCal towers to a reference shower shape 
expected for electromagnetic showers. Most hadrons are not absorbed in the 
EMCal and traverse it as minimum ionizing particles. The typical hadron 
energy loss in the EMCal is $\sim$ 0.3\,GeV~\cite{em_profile_cut}. To 
reduce hadron contamination and to account for the poorer EMCal resolution 
at lower energies, a minimum energy $E_{\g}>0.2$\,GeV is 
required for clusters reconstructed in all \dau events and in peripheral 
\cucu events. In more central \cucu collisions it is increased to 
$E_{\g}>0.4$\,GeV. The two clusters from the same \piz meson are also 
required to fall within the acceptance of the same EMCal sector  
to suppress boundary effects. The energy balance between the two clusters 
forming a \piz candidate is characterized by $\alpha=|E_1-E_2|/|E_1 + 
E_2|$, where $E_1$ and $E_2$ are the cluster energies. For 
$\piz\rightarrow \g\g$ decays the parameter $\alpha$ has an almost flat 
distribution between 0 and 1~\cite{em_profile_cut} . Due to the steeply 
falling \pt spectrum of all particles produced in the event, most of the 
EMCal clusters have a low energy partner, therefore the distribution of 
the parameter $\alpha$ calculated for combinatorial pairs has a distinct 
peak close to 1 for high \pt pairs. To exclude those pairs, parameter 
$\alpha$ is required to be less than 0.8.

A pair of \g-clusters is selected as a \piz candidate if its reconstructed 
invariant mass is within $\pm2$ standard deviations from a parameterized 
\piz mass:
\begin{eqnarray}
\label{pizenergycorr}
|M_{\g\g}(\pt) - M_{\piz}(\pt) \times R_M(\pt)|&<&2 \sigma_{\piz}(\pt)  \nonumber \\
                                              & \times & R_\sigma(\pt),
\label{eq:widths}
\end{eqnarray}
where $M_{\g\g}$ is the reconstructed invariant mass of a pair of the 
\g-clusters, \pt is the transverse momentum of the pair, $M_{\piz}(\pt)$ 
and $\sigma_{\piz}(\pt)$ are the parameterizations of the mass and 
1-$\sigma$ width of the \piz peak as a function of transverse momentum.  
The parameterization is performed using an inclusive sample of \piz 
mesons. $R_M(p_T)$ and $R_\sigma(p_T)$ are correction factors accounting 
for the difference between inclusive \piz mesons and neutral pions produced in 
\kshort meson decays.

To determine $M_{\piz}(\pt)$ and $\sigma_{\piz}(\pt)$, the peak position 
and width of the \piz peak in the invariant mass distribution of the 
cluster pairs are measured for different \pt bins and are parameterized as 
a function of \pt. The mass and width of \piz are determined by fitting 
the invariant mass distribution with a sum of a Gaussian function 
describing the signal and a second order polynomial describing the 
background. Figure~\ref{fig:ksmasswidth} shows reconstructed mass and 
width of \piz as a function of \pt in \cucu collisions for one of the 
EMCal sectors.

Because of the long lifetime of the \kshort meson, the neutral pions from 
its decay are produced at a displaced vertex and thus the momentum 
components of the clusters are mis-reconstructed. This results in a 
different reconstructed mass and width of \piz mesons from \kshort decays 
compared to those reconstructed for inclusive \piz mesons that mostly 
originate from the event vertex. In the data we have no means to isolate a 
sample of neutral pions from \kshort meson decays. Therefore a 
quantitative study of this effect is possible only in Monte Carlo 
simulation.  Samples of \piz mesons produced from the decay of \kshort 
mesons with a realistic \pt distribution and neutral pions produced at the 
primary collision vertex with the inclusive \pt distribution were 
generated. Neutral pions were reconstructed using the same analysis chain 
as in real data. From Fig.~\ref{fig:ksmasswidth}~(a) and (b), one can see 
the reconstructed masses and widths of simulated inclusive \piz mesons 
(circles) originating from the event vertex are consistent with the values 
measured in real data (open crosses). Neutral pions from \kshort decays 
are reconstructed with smaller mass and larger width. The correction 
factors $R_M(p_T)$ and $R_\sigma(p_T)$ are calculated as the ratio of the 
parameterizations of $M_{\piz}(\pt)$ and $\sigma_{\piz}(\pt)$ for neutral 
pions from \kshort mesons and inclusive \piz mesons. These correction 
factors improve the signal-to-background ratio by 30\%--50\%.

The \kshort mesons are reconstructed by combining the \piz candidates in 
pairs within the same event. Pairs of \piz candidates that share the same 
cluster are rejected. To improve the signal-to-background ratio \piz 
candidates are required to have $p_T>1.0$\,\gevc in the \dau sample 
and $p_T>1.5$\,\gevc for \cucu events with centrality 
$>20$\% and $p_T>2$\,\gevc for \cucu events with 
centrality $<20$\%.

The red squares in Fig.~\ref{fig:ksinvmass_comp} give an example of the 
invariant mass distribution for $\piz\piz$ pairs measured in the minimum 
bias \dau collisions at $8<p_T<9$~GeV/$c$. Due to the steeply falling \pt 
spectrum of produced particles, the finite energy/position resolution and 
nonlinear response of the EMCal, the reconstructed mass of \piz mesons 
differs from the nominal PDG value $M_{PDG}=134.98$\,MeV~\cite{PDG}. To 
match the reconstructed mass of \piz candidates to the PDG value, the 
energy and momentum of clusters building a pair are multiplied by the 
ratio of measured and nominal \piz mass: $M_{PDG}/M_{\g\g}$. This 
correction decreases the width of reconstructed \kshort meson peak by 
$\approx\,50$\%.  An example of the invariant mass distribution after 
energy correction is shown with blue open crosses in 
Fig.~\ref{fig:ksinvmass_comp}.  The black circles correspond 
to the case when \piz candidate selection is changed according to 
Eq.~\ref{pizenergycorr} to account for the difference between inclusive 
\piz mesons and neutral pions produced in \kshort meson decays.

\begin{figure*}[htb]
  \includegraphics[width=0.45\linewidth]{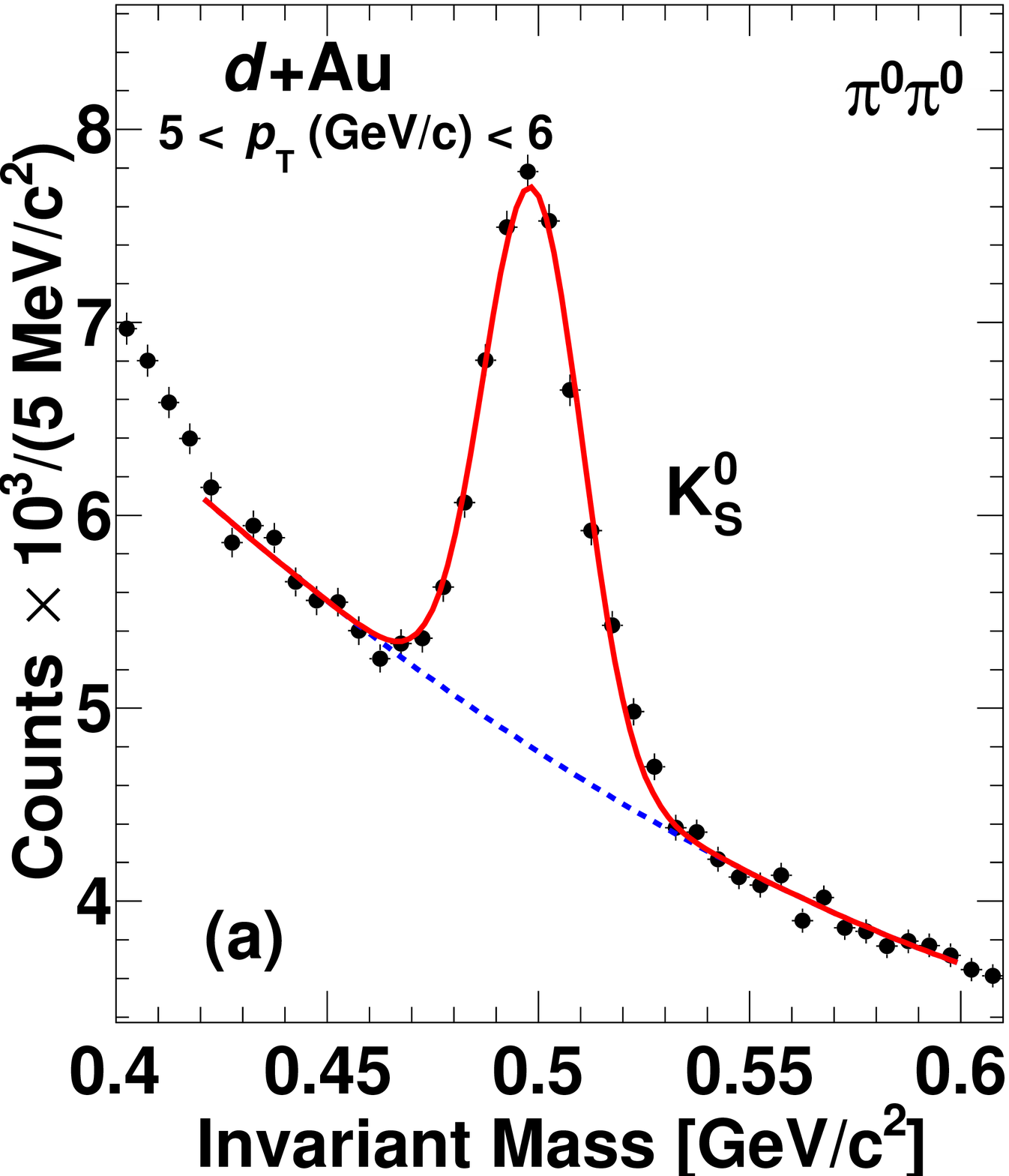} 
  \includegraphics[width=0.45\linewidth]{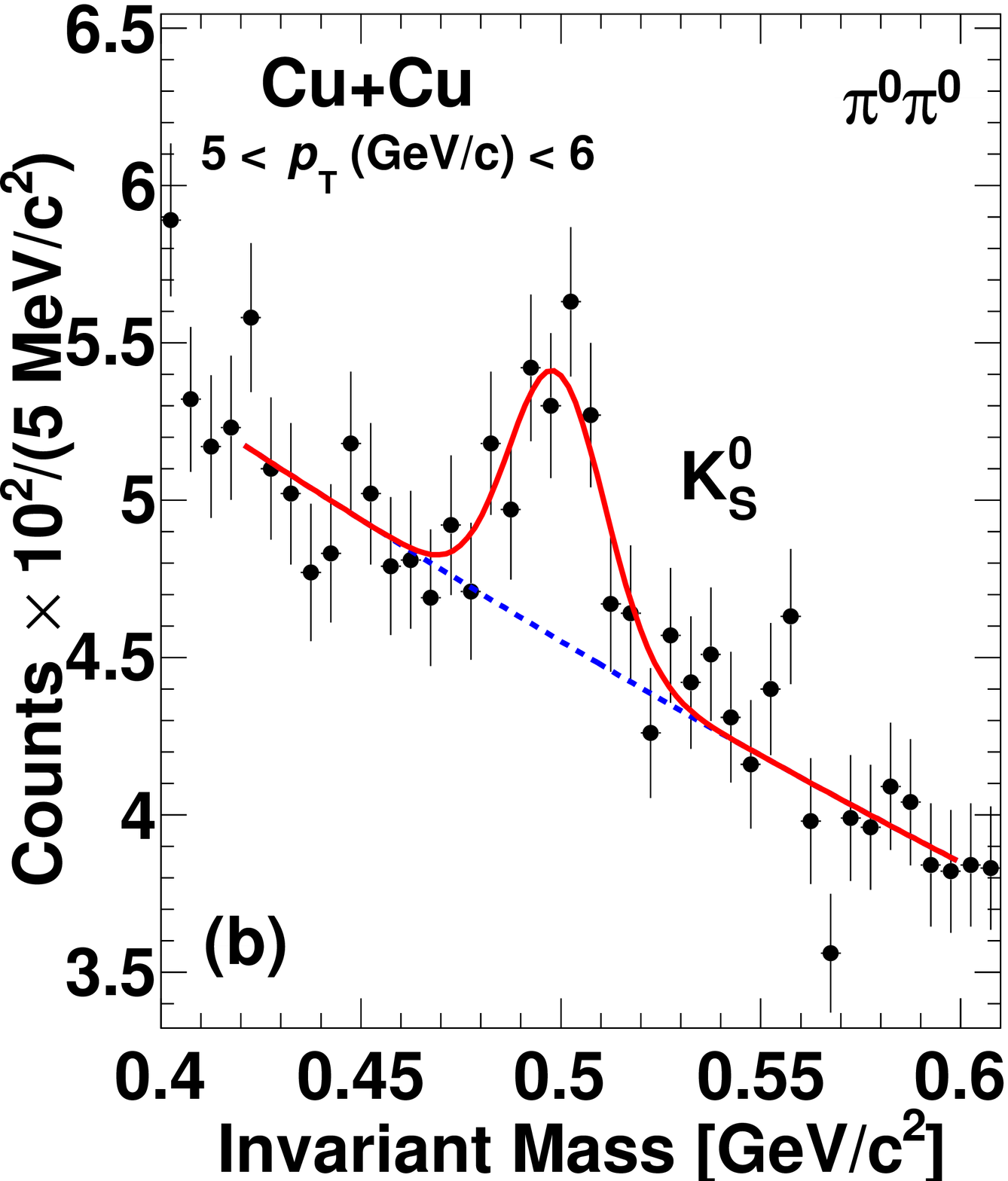} 
  \caption{(color online)
The invariant mass reconstructed from two \piz mesons in the range  
$5<p_T<6$\,\gevc in (a) \dau and (b) \cucu collisions at 
\energy for the MB data. The distributions are approximated by a Gaussian 
plus a second order polynomial shown by solid red and blue dashed lines 
respectively.
}
\label{fig:ks_invmassplot} 
\end{figure*}

The \kshort meson raw yield in each \pt bin is extracted by fitting the 
$\piz\piz$ invariant mass distribution to a combination of a Gaussian 
function for the signal and a polynomial for the background. A second 
order polynomial provided adequate description of the background shape 
outside of the \kshort peak and varied smoothly under the peak. The 
fitting range was set to about $\pm8$ standard deviations from the peak 
center and was enough to constrain the fit. A wider fitting range would 
require a higher order polynomial to describe the background. All fits 
resulted in $\chi^2$/NDF values close to one. The \kshort meson yield in 
each \pt bin is calculated as the integral of the Gaussian function. 
Examples of $\piz\piz$ invariant mass distributions are shown in 
Fig.~\ref{fig:ks_invmassplot}~(a) and (b) for \dau and \cucu, 
respectively.

The typical signal/background ratio, integrated within $\pm$2$\sigma$ 
around particle mass, for different centrality classes grows from 0.5 to 
0.86 (0.04--0.85) in \dau (\cucu) collisions with increasing transverse 
momentum. The width and the mass of the reconstructed \kshort mesons were 
found to be in good agreement with the values expected from simulation.

\subsection{Reconstruction of \kstar meson invariant mass \label{kstar_ana}}

The \kstar and \kstarbar mesons are reconstructed from their hadronic 
decay channels $K^{+}\pi^{-}$ and $K^{-}\pi^{+}$, respectively. We denote 
the average of \kstar and \kstarbar as \kstar. Tracks selected for this 
analysis are required to have $p_T>0.3$\,\gevc. The TOF system 
covers approximately one half of the east central arm spectrometer 
acceptance and can identify charged kaons up to approximately 
2.5\,\gevc~\cite{TOF_ppg101}. To extend the high \pT reach of the \kstar 
meson measurement, unidentified, oppositely charged tracks are also 
included in the analysis. These tracks are required to have associated 
hits in PC3 or EMCal and are referred to as the PC3-matched tracks. 
Depending on the track selection criteria, three different techniques are 
considered in this analysis.

\begin{enumerate}

\item \textit{fully identified} where tracks are identified as kaon and 
pion in TOF.

\item \textit{kaon identified} where one of the tracks is identified as 
kaon in TOF and the other is a PC3-matched track to which the pion mass is 
assigned.

\item \textit{unidentified} where both tracks are the PC3-matched tracks.

\end{enumerate}

The three techniques are exclusive to each other and statistically 
independent. The PC3-matched tracks are assigned the nominal mass of the 
$\pi$ or $K$ mesons depending on which technique is used. The \pt ranges 
accessible in the different techniques in \pp, \dau and \cucu collisions 
are given in Table~\ref{sampletype}.

\begin{table}
\caption[] {Different techniques  used in \kstar measurement and their \pt coverage
in \pp, \dau and \cucu collisions at \energy. 
The table also shows the range of signal-to-background, integrated within $\pm$3$\sigma$ 
around particle mass (S/B),  values for each sample.}
\label{sampletype}
\begin{ruledtabular}\begin{tabular}{cccc}
Collision         & Technique             & \pt range   &   S/B \\
System           & used                       &   (GeV/$c$)          & \\
\hline
\pp               & fully identified   &  1.1--4.0  &   0.011--0.023 \\
                  & kaon identified   &  1.1--4.0  &    0.005--0.0147 \\
                  & unidentified        &  2.3--8.0  &  0.006--0.021 \\
\dau              & fully identified   &  1.1--4.0  &   0.009--0.015\\
                  & kaon identified   &  1.4--4.5  &    0.003--0.0118\\
                  & unidentified        &  2.3--8.5  &  0.009--0.012\\
\cucu             & fully identified   &  1.4--4.0  &   0.0048--0.0076 \\
                  & kaon identified   &  1.7--4.5  &    0.0006--0.0039 \\
                  & unidentified        &  2.9--8.0  &  0.0011--0.0036 \\
\end{tabular}\end{ruledtabular}
\end{table}

The ``fully-identified'' sample with both charged particles identified in 
the TOF has the highest signal-to-background ratio and provides access to 
\kstar meson production at low and intermediate \pt. However, due to the 
limited PID capabilities of the TOF technique and the small acceptance of 
the TOF detector, this data set does not provide sufficient statistical 
precession for $p_T>4$\,\gevc. The ``kaon identified'' sample 
allows for the best signal extraction at intermediate \pt. The 
``unidentified'' sample has a poor signal-to-background ratio that 
prevents signal extraction at low \pt. Signal extraction is possible at 
higher $p_T>2.3$\,\gevc in \pp or \dau collisions and 
$p_T>2.9$\,\gevc in \cucu collisions), because of the smaller 
combinatorial background. The highest \pt reach of \kstar measurements 
with the ``unidentified'' sample is limited only by the sampled 
luminosity. Measurements performed with the three techniques have a wide 
overlap region that is used for evaluation of the systematic 
uncertainties.

\begin{figure*}
  \includegraphics[width=0.32\linewidth]{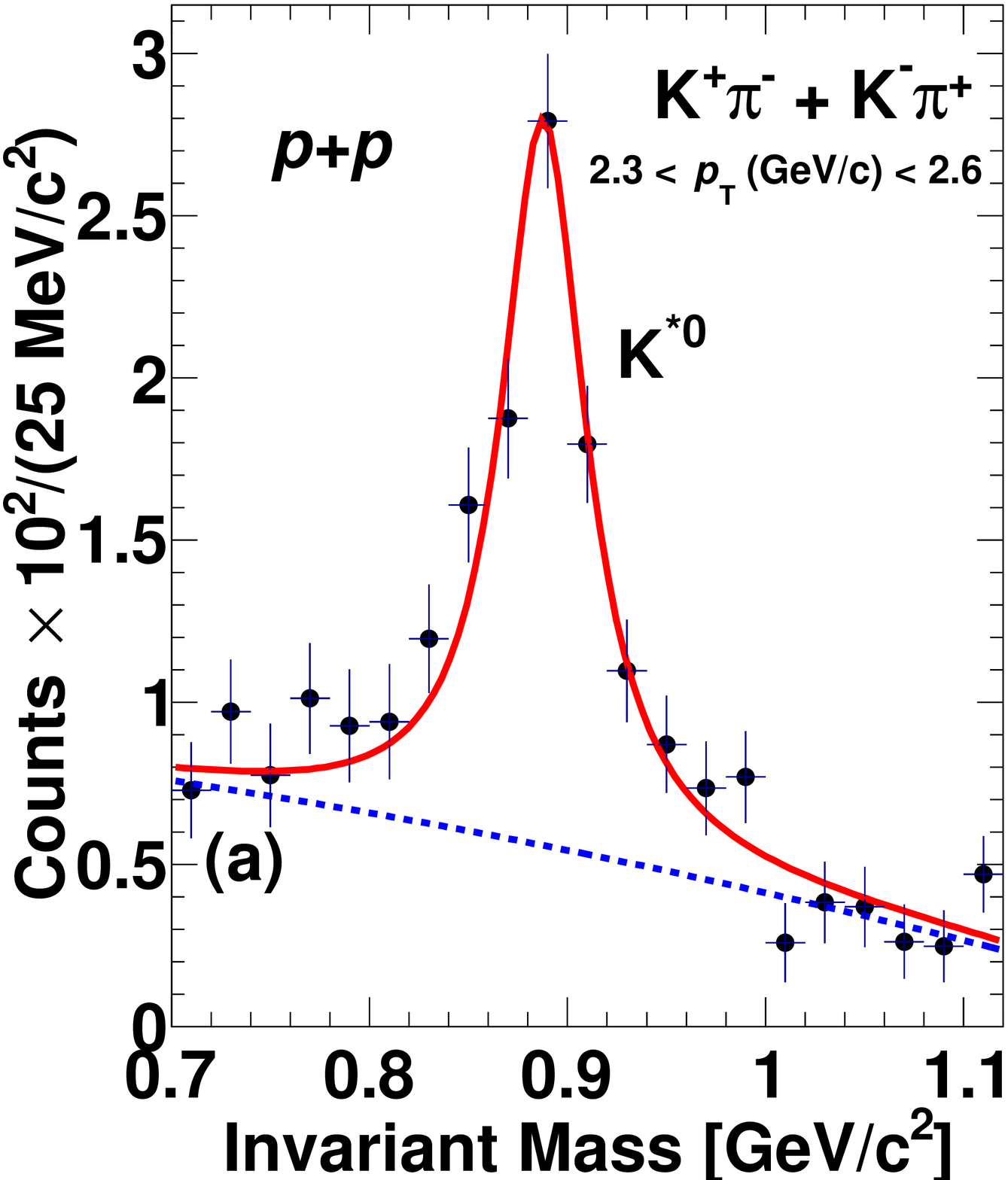}
  \includegraphics[width=0.32\linewidth]{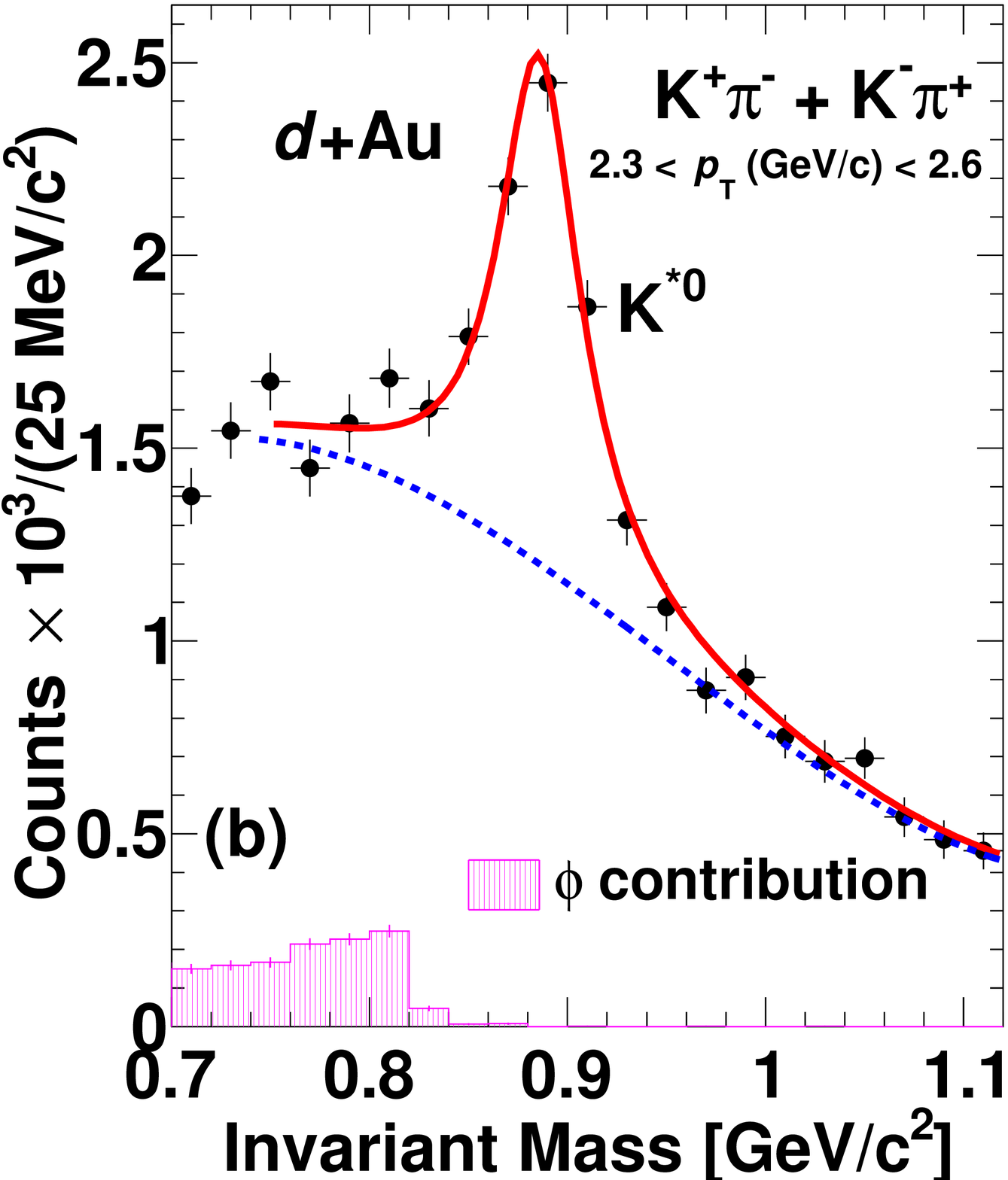}
  \includegraphics[width=0.32\linewidth]{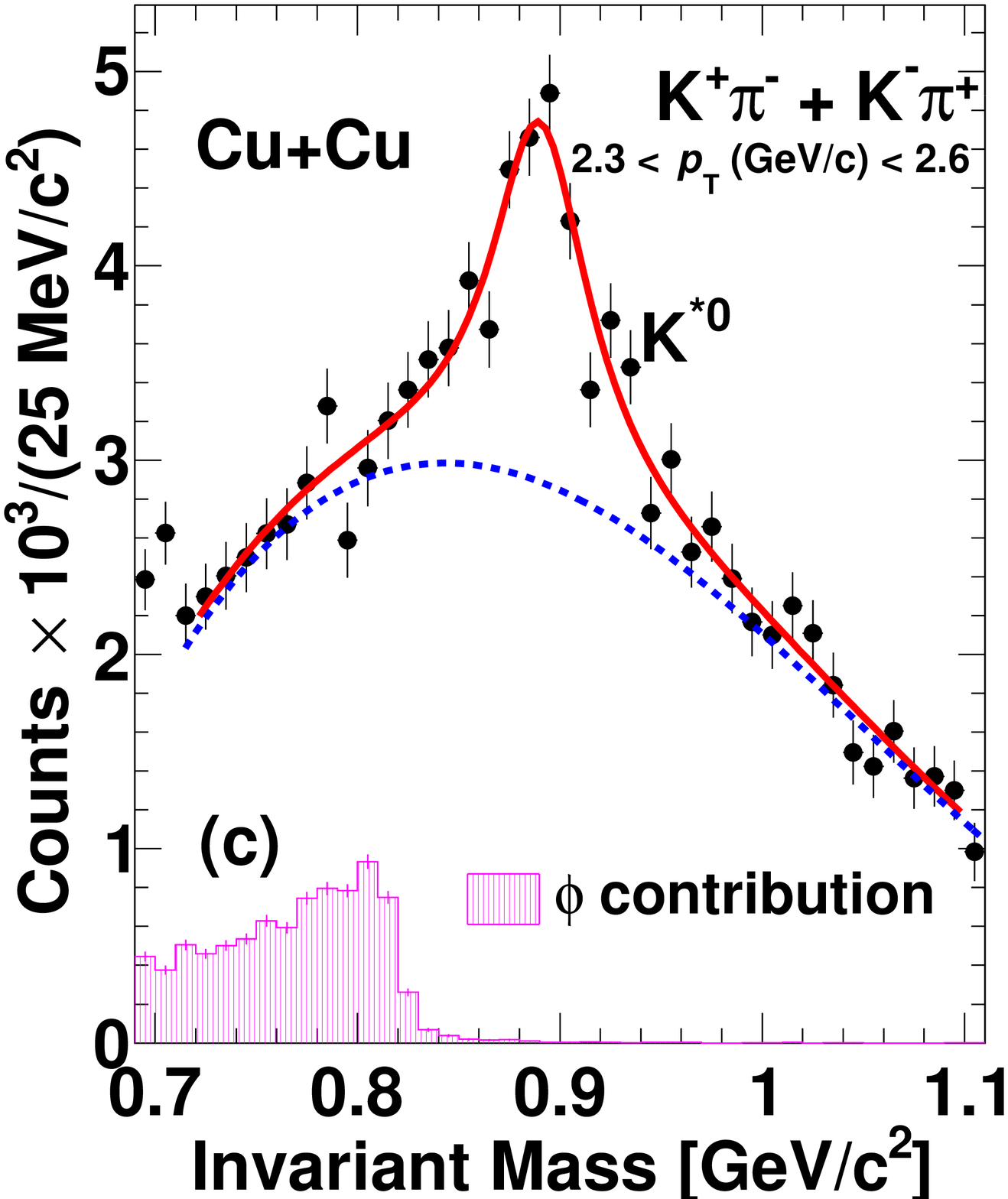}
\caption{(color online)
The invariant mass distributions of \kpi candidates, where $K$ is 
identified in TOF and $\pi$ is matched in PC3, in the range 
$2.3<p_T<2.6$\,\gevc for (a) \pp, (b) \dau, and (c) \cucu collisions at 
\energy.  The distributions are shown after subtraction of the mixed event 
background and the correlated background from misidentified $\phi 
\rightarrow K^+K^-$ and \kshort$\rightarrow$$\pi^{+}\pi^{-}$ decays (see 
text for details). The distributions are fitted to the sum of the RBW 
function for the signal and a polynomial (second order in \pp and third 
elsewhere) for the background shown with solid red line. The residual 
background is also shown separately with blue dashed line. The $\phi$ 
contribution is shown by the magenta colored histogram.
}
\label{fig:kstarpeaks}
\end{figure*}

The invariant mass distribution for \kpi pairs comprises both signal and 
background. The uncorrelated part of the background that arises from the 
random combination of tracks in the same event is estimated using the 
mixed event technique~\cite{PPG016}. The event mixing combines positively 
(negatively) charged tracks from one event with the charged tracks of 
opposite sign from another event within the same centrality class. The 
number of mixed events for each event in the data is set to 20 for \pp and 
\dau and to 10 for \cucu collisions, to have sufficient statistics. The 
mixed event invariant mass distribution is normalized by the number of 
events mixed and then it is subtracted from the unlike sign distributions. 
The correlated part of the background is dominated by track pairs from 
mis-reconstructed or not fully reconstructed decays of light hadrons. Two 
such processes, $\phi \rightarrow K^{+}K^{-}$ and 
$\kshort\rightarrow\pi^{+}\pi^{-}$, produce smeared peak structures in the 
invariant mass distribution in the close vicinity of the \kstar mass peak. 
Contributions of these two sources are estimated using measured yields of 
the $\phi$ meson~\cite{phipaper} and \kshort meson~\cite{PPG099}. The 
location and shape of these peaks are modeled by the PHENIX based 
simulations. The estimated contributions are then normalized by the number 
of events analyzed for \kstar meson and subtracted from the measured 
\kstar invariant mass distributions. Apart from these contributions, a 
residual background due to other correlated sources~\cite{starKstarpp} 
remains in the subtracted spectra. The residual background is different 
depending on the collision systems, analysis techniques and also on the 
pair \pt. Examples of invariant mass distributions after subtraction of 
the mixed event background and the correlated background from \kshort and 
$\phi$ mesons are shown in Fig.~\ref{fig:kstarpeaks}~(a), (b) and (c) for 
\pp, \dau and \cucu collisions, respectively. The $\phi$ contribution is 
shown by the magenta colored histogram. It is seen that this contribution 
is very small in \cucu case, even smaller in \dau case and negligible in 
\pp case. The residual background is clearly seen in the subtracted mass 
spectra. In the ``fully-identified technique'', this residual background 
is relatively small. It is larger in the ``kaon-identified technique'' and 
even larger in the analysis based on unidentified tracks.

The invariant mass distribution in each \pt bin is fit to the sum of a 
relativistic Breit-Wigner (RBW) function for the signal and a 
$2^{\rm nd}$ or $3^{\rm rd}$ order polynomial for the residual background.
\begin{eqnarray}
  RBW = \frac{1}{2\pi} \frac{M_{K\pi}M_{K^{*0}}\Gamma}{(M_{K\pi}^{2} - M_{K^{*0}}^{2})^{2} + 
M_{K^{*0}}^{2}\Gamma^{2}},
\end{eqnarray} 
where $M_{K\pi}$ is the reconstructed invariant mass, $M_{K^{*0}}$ is the 
fitted mass of \kstar meson and $\Gamma$ is the width of \kstar meson 
fixed to the value obtained from simulation. Because the experimental mass 
resolution ($\sim$ 5\,MeV/$c^2$) is much smaller than the natural width of 
the \kstar meson the simulated $\Gamma$ is very close to the nominal width 
of 48.7\,MeV/$c^2$~\cite{PDG}.

The raw yield of the \kstar meson in each \pt bin is obtained as follows. 
The yield in each \pt bin is summed up in the invariant mass window of 
$\pm$ 75 MeV/$c^2$ around the nominal mass of \kstar meson which includes 
both signal and residual background. The invariant mass distribution is 
fitted, as explained above and the residual background contribution is 
obtained by integrating the background component of the fit (second or 
third order polynomial) in the same mass window. The residual background 
contribution is subtracted from the total signal to obtain the raw yield 
for \kstar meson.

\subsection{Calculation of invariant yield}
\label{inv_yields}

The invariant yields of \kshort and \kstar mesons are calculated by
\begin{eqnarray}
\label{formula2}
\frac{1}{2 \pi p_T} \frac{d^{2}N}{dp_Tdy} & = &\frac{1}{2\pi p_T \, \Delta p_T \, \Delta y} 
\nonumber \\
& \times &   \frac{Y_{\rm raw}}{N_{\rm evt} \, \epsilon  (\pt)\, BR} \, \times 
\frac{C_{bias}}{\epsilon_{treff}},
\end{eqnarray}
where $Y_{\rm raw}$ is the meson raw yield (see Sections~\ref{kshort_ana} 
and \ref{kstar_ana}), $N_{\rm evt}$ is the number of sampled events in the 
centrality bin and $\epsilon(\pt)$ includes geometrical acceptance, 
reconstruction efficiency, and occupancy effects in the high multiplicity 
environment of heavy ion collisions. The branching ratio ($BR$) for 
$\kshort\rightarrow\piz\piz$ is 30.69 $\pm$ 0.05\% (BR for $\piz 
\rightarrow 2\gamma$ is $98.823\pm0.034$\%). The branching ratio for the 
\kstar $\rightarrow K^{+}\pi^{-}$ is close to 67\%. The trigger bias 
correction $C_{bias}$ is 0.69~\cite{phipaper} for \pp collisions and for 
\dau collisions it varies from 1.03 to 0.94~\cite{PPG146} with increasing 
centrality. The trigger bias correction in \cucu collision system is taken 
equal to unity in all analyzed centrality bins. The ERT efficiency 
for \kshort meson $\epsilon_{treff}$ determines the probability of 
$\kshort\rightarrow\piz\piz \rightarrow 4\gamma$ decay products to fire 
the ERT. For the \kstar which uses no additional trigger, 
$\epsilon_{treff}=1$.

The invariant cross section in the \pp system is given by :
\begin{eqnarray}
\label{formula1}
 E \frac{d^{3}\sigma}{dp^{3}} = \sigma_{pp}^{inel} \times 
\frac{1}{2\pi\pt} \frac{d^{2}N}{d\pt dy},
\end{eqnarray}
where $\sigma_{pp}^{inel}=42.2\pm3$\,mb~\cite{PPG099} is the total 
inelastic cross section in \pp collisions at \sqs = 200\,GeV.

\begin{figure*}[htb]
  \includegraphics[width=0.45\linewidth]{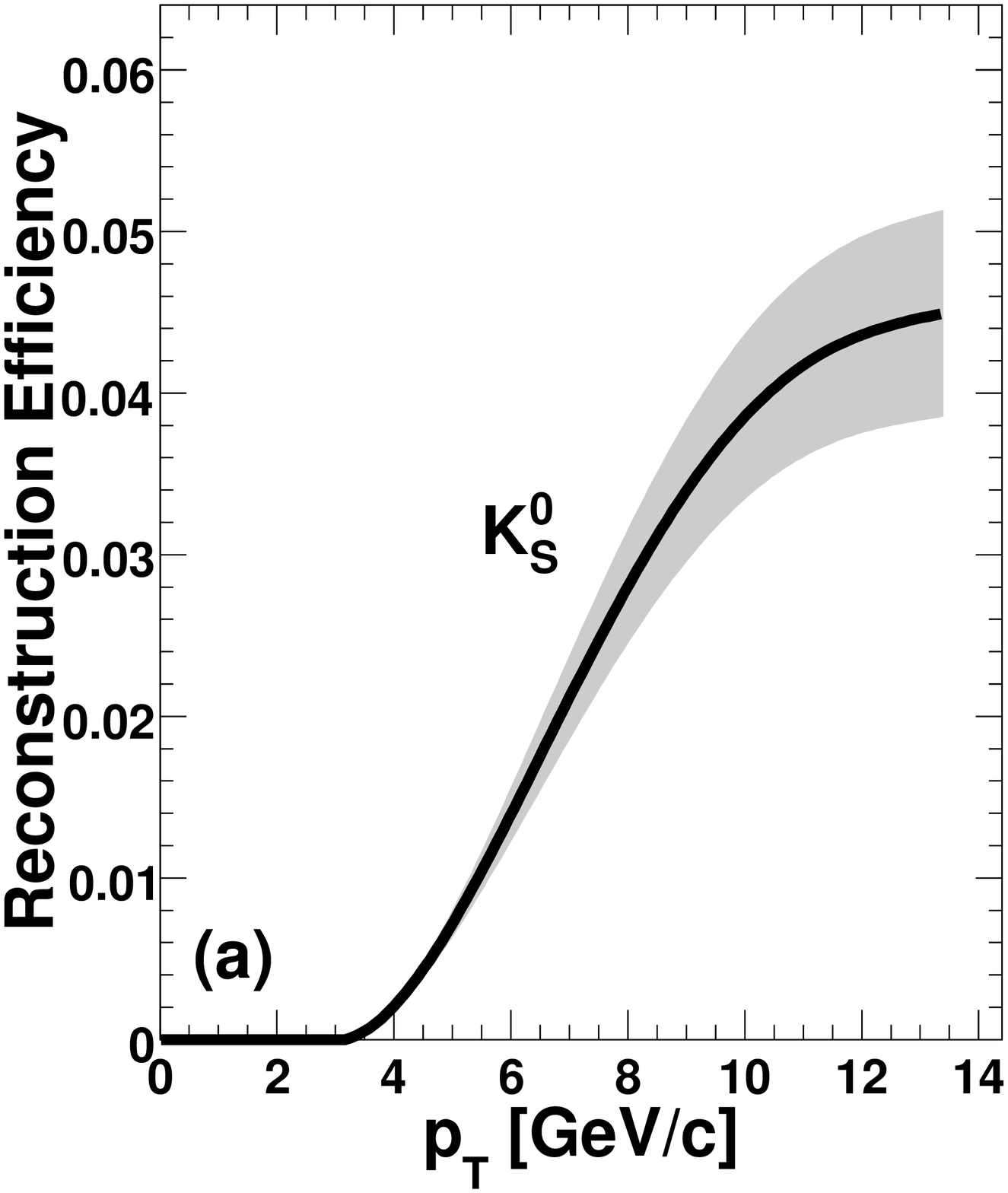}
  \includegraphics[width=0.45\linewidth]{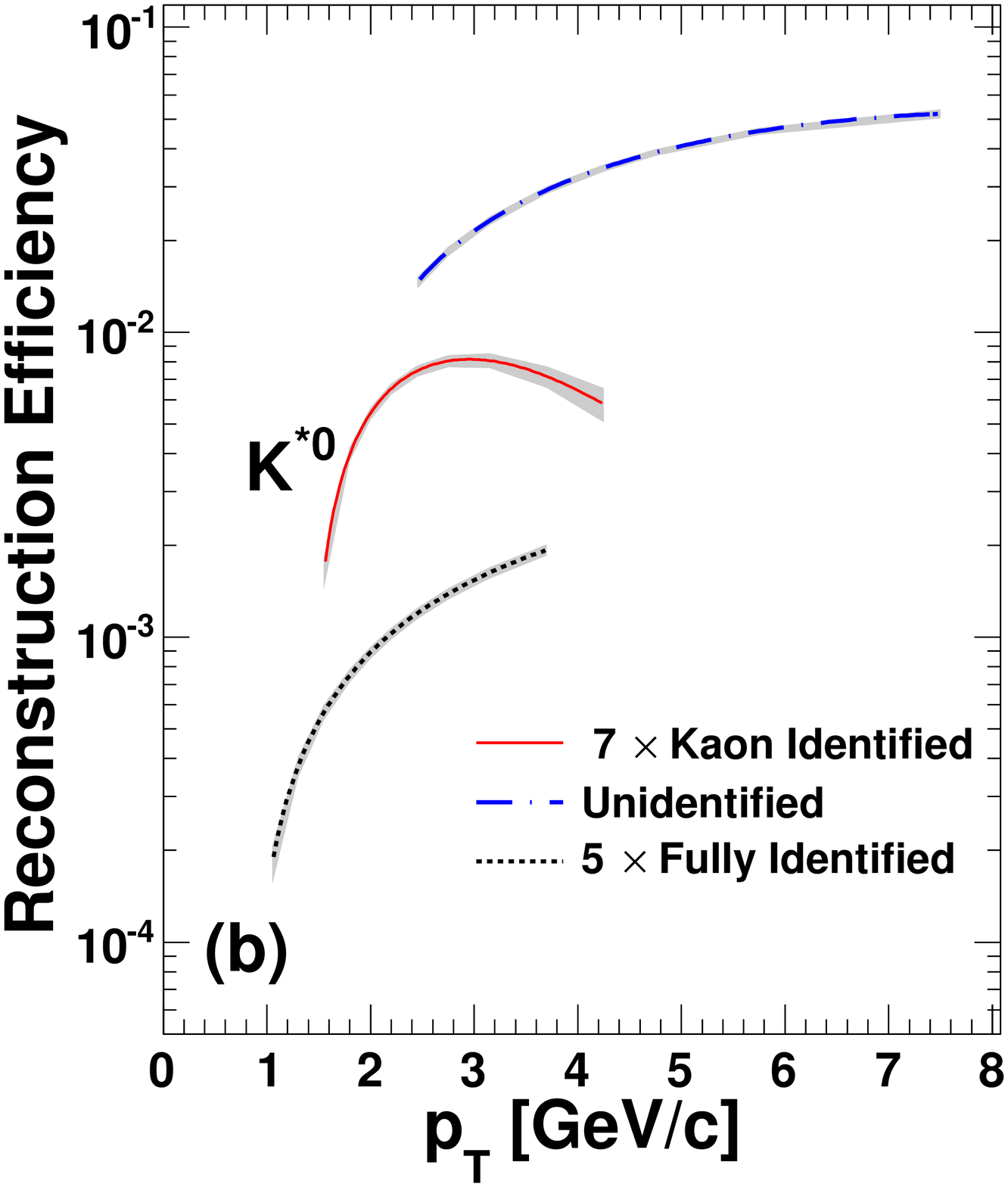}
  \caption{(color online)
Reconstruction efficiency for (a) \kshort and (b) \kstar for \dau 
collisions. The gray band shows the systematic uncertainty. Please refer 
to Table~\ref{table_syserr_kshort} for systematic uncertainties. Fig.~(b) 
shows the reconstruction efficiency for the ``kaon identified'', 
``unidentified'' and ``fully identified'' techniques for \kstar analysis 
are shown by the red solid line, dotted dashed blue line and black dashed 
line, respectively.
}
  \label{fig:eff} 
\end{figure*}

The reconstruction efficiency for the \kshort and \kstar mesons are 
obtained from Monte Carlo simulations. Both the \kshort and \kstar mesons 
are generated using single particle event generator Exodus~\cite{PPG088}. 
The primary mesons are decayed into the measured channel and all particles 
are traced through the PHENIX setup using the {\sc geant}~\cite{GEANT} based 
PHENIX simulation package. The decayed particles are reconstructed using 
the same analysis procedures as used in the analysis of real data. The 
reconstruction efficiency is calculated as the ratio of the number of 
reconstructed mesons counted in the same way as in data, to the number of 
generated mesons. Due to high detector occupancy in \cucu collisions, the 
reconstruction efficiency becomes smaller due to hit and cluster merging 
in detector subsystems. To take this effect into account the 
reconstruction efficiencies for \kshort and \kstar mesons were determined 
after embedding the simulated signals in real events. The \kstar meson 
reconstruction efficiency in \cucu is reduced by $\sim$ 5\% in the most 
central collisions and by $\sim$ 1\% in peripheral collisions. These 
corrections are included in $\epsilon(\pt)$, as shown in 
Fig.~\ref{fig:eff}.

\begin{figure*}[htb]
    \includegraphics[width=0.48\linewidth]{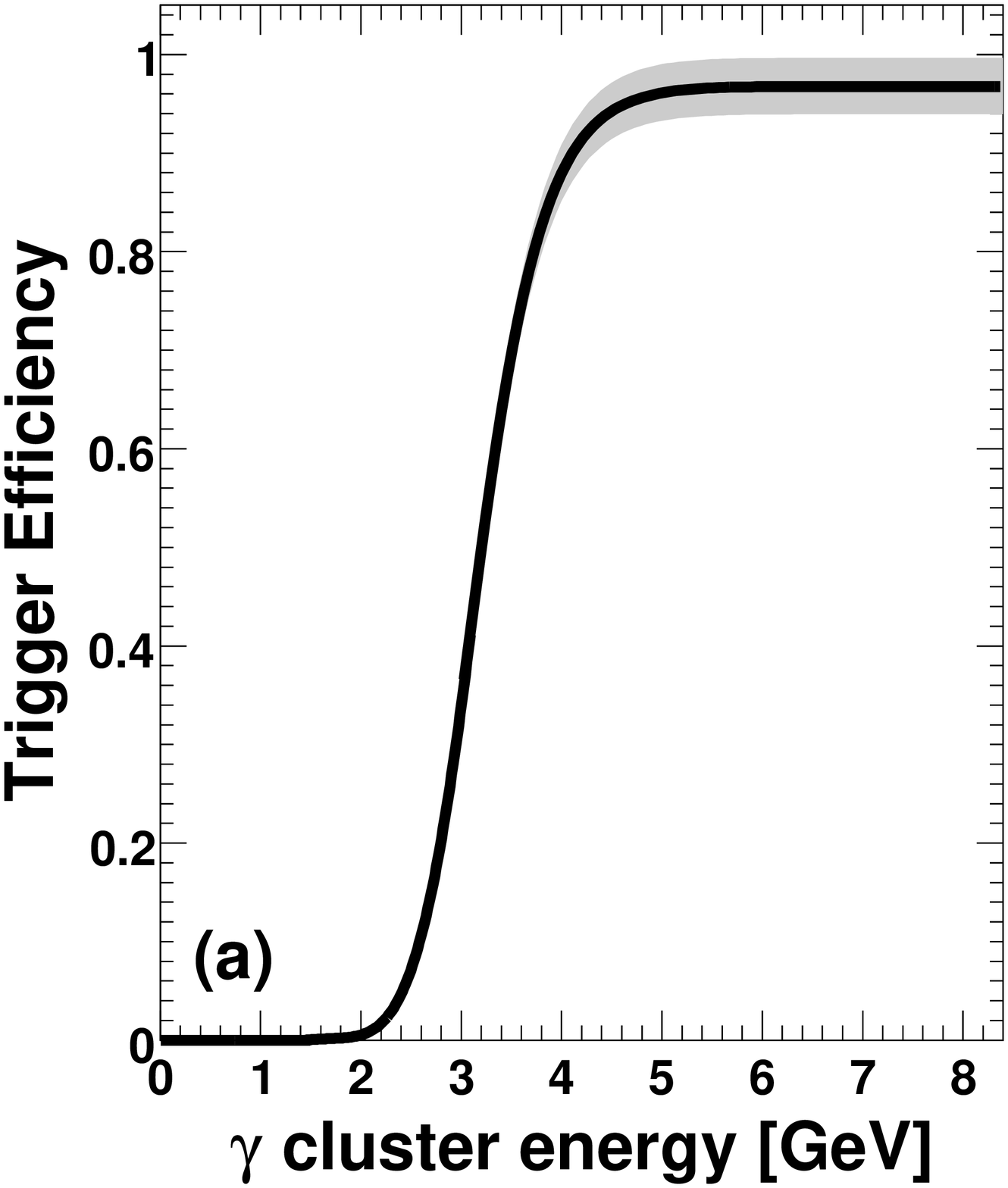}
    \includegraphics[width=0.48\linewidth]{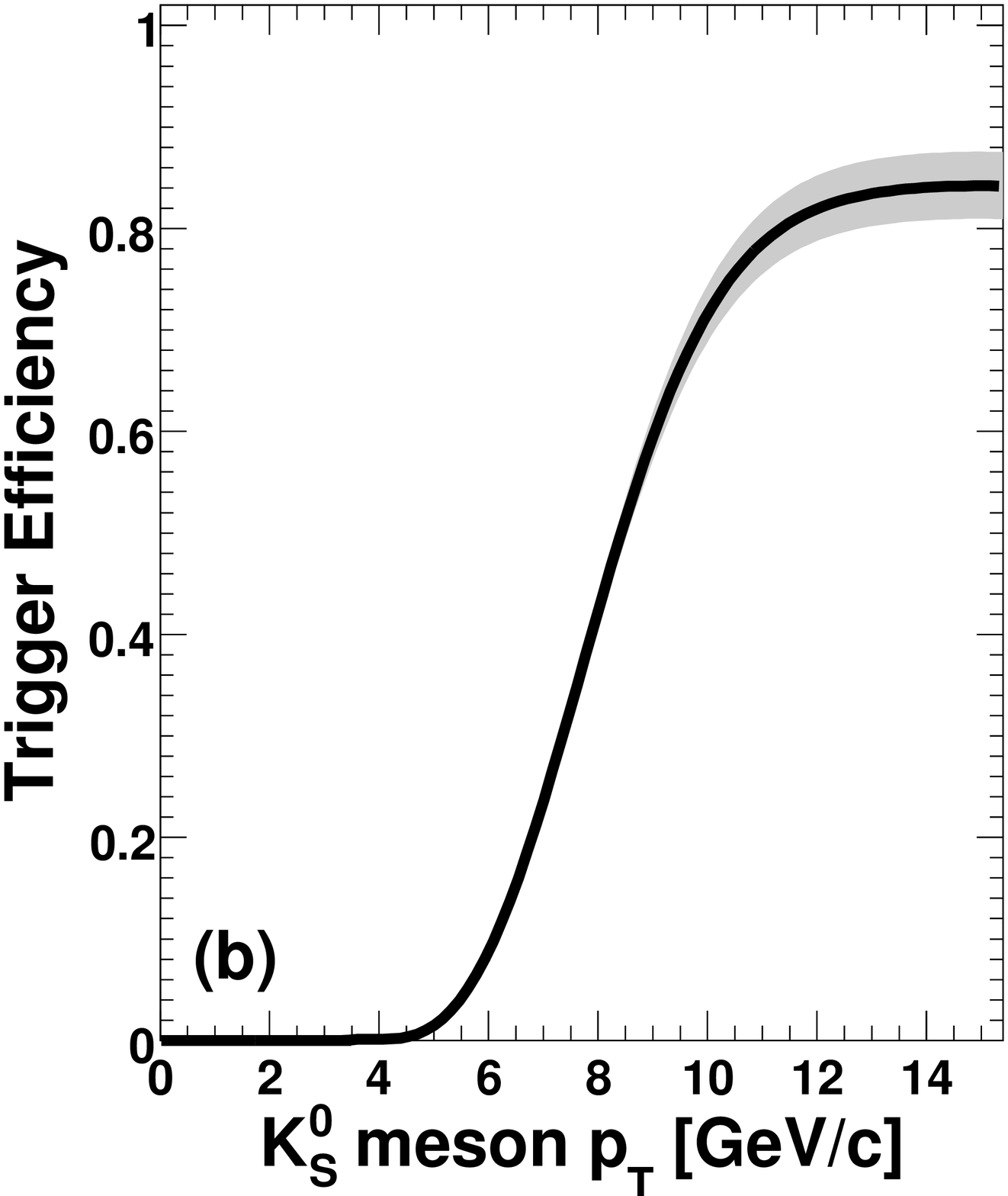}
\caption{
(a) Trigger efficiency for single photons as a function of 
cluster energy. (b) \kshort trigger efficiency as a function of \pt. The 
bands show the systematic uncertainty. Results are presented for the \cucu 
data recorded in 2005.
}
    \label{fig:ks_ert} 
\end{figure*}

The probability that one of the \kshort meson decay products fires the ERT 
trigger is estimated based on the measured single photon ERT 
efficiency, $\epsilon_{\gamma}$. The latter is evaluated as the ratio of 
the number of clusters that fired the ERT to the number of 
clusters of the same energy in the minimum bias data sample. The trigger 
efficiency is calculated as a function of cluster energy separately for 
each EMCal sector. An example of $\epsilon_{\gamma}$ in one of the EMCal 
sectors is shown in Fig.~\ref{fig:ks_ert}~(a) for the case of 2005 
measurements for \cucu collisions.

The trigger efficiency grows steeply with energy and reaches 50\% at the 
energy approximately corresponding to the ERT threshold setting. 
The curves saturate at approximately twice the threshold energy. The level 
of saturation is below 100\% because of inactive areas of the ERT. The 
trigger efficiency for \kshort meson ($\epsilon_{treff}$) is evaluated 
using Monte Carlo simulation. The \kshort meson is considered to fire the 
ERT if at least one of the photons in the final state fires the 
trigger. The resulting trigger efficiency for \ksdecay is shown in 
Fig.~\ref{fig:ks_ert}~(b). The trigger efficiency uncertainty for \kshort 
meson was evaluated by varying the single photon ERT efficiency 
within the uncertainties of the measurement.

\subsection{Systematic Uncertainties}
\label{sys_errs}

Several factors contribute to the systematic uncertainty of the 
measurement of the \kshort meson invariant yield: the raw yield 
extraction, the reconstruction efficiency and detector acceptance and the 
$\kshort\rightarrow\piz\piz$ decay branching ratio uncertainty. Evaluation 
of the systematic uncertainties associated with the \kshort meson raw 
yield extraction is done by varying the raw yield extraction method and by 
modifying the background shape around the \kshort peak. The $\piz\piz$ 
invariant mass distribution is approximated by a second order polynomial 
outside three standard deviations from the center of the peak region. The 
polynomial is then interpolated under the peak and subtracted from it. The 
yield is obtained by integrating the subtracted invariant mass 
distribution in a three standard deviation window around the mean of the 
peak. To modify the background shape the ``cross \piz meson'' cut is used. 
This cut significantly changes the background shape in the invariant mass 
distributions of $\piz\piz$ pairs in the vicinity of the \kshort meson 
peak. If two photons with the largest energy, assigned to different \piz 
candidates, produce an invariant mass within ${\pm}4\times\sigma_{\pi_0}(p_T)$ 
from the $M_{\piz}(\pt)$ given in Eq.~\ref{eq:widths}, the entire 
combination of four clusters is rejected. The RMS of the corrected raw 
yields obtained in all combinations of yield extraction and background 
modification is taken as an estimate of the systematic uncertainty for the 
signal extraction.

The uncertainty in the reconstruction efficiency is dominated by 
mismatches in detector performance between data and Monte Carlo. The 
uncertainty on the EMCal acceptance is estimated by artificially 
increasing dead areas in the EMCal by 10\% and redoing the analysis. To 
estimate the contribution of the EMCal energy resolution to the systematic 
uncertainty, the \kshort meson reconstruction efficiency is recalculated 
with the energy resolution artificially worsen by 3\%. The 3\% variation 
of the energy resolution was chosen as a maximum value that would still 
provide consistency between the \piz meson widths from real data and 
simulations. The contribution of the EMCal energy scale uncertainty was 
estimated by varying the energy scale within $\pm$1\% in simulation. The 
variation range is constrained by the \piz meson peak positions in real 
data and simulation. Photon conversion in the detector material is 
accounted for in the calculation of the reconstruction efficiency. 
However, detector materials are described in the simulation with some 
precision and thus an uncertainty associated with the photon conversion is 
introduced. The conversion correction uncertainty was estimated in 
Ref.~\cite{em_profile_cut} to be equal to 3\% for the neutral pions. Thus 
the \kshort meson conversion correction uncertainty is 6\%.

The \piz meson candidates are selected within two standard deviations 
around the \piz meson peak position in the invariant mass distribution of 
two photons. The difference between the \piz meson width parameterizations 
in real data and Monte Carlo simulations does not exceed 10\%. To estimate 
the \piz selection cut uncertainty, the window around the \piz meson peak 
position is varied by 10\%. The difference between the \kshort meson 
reconstruction efficiencies calculated with changed and default cuts is 
taken as the uncertainty related to the \piz candidate selection cut. The 
\kshort meson trigger efficiency uncertainty is evaluated by varying the 
single photon $\epsilon_{\g}$ trigger efficiency within uncertainties of 
its measurement. Relative systematic uncertainties for the \kshort meson 
measurements in \dau and \cucu systems are given in 
Table~\ref{table_syserr_kshort}. The uncertainties are categorized by 
types: A, B and C. Type A denotes the \pt uncorrelated uncertainty, type B 
denotes the \pt correlated uncertainty and type C denotes the overall 
normalization uncertainty such as the minimum bias trigger efficiency in 
\pp and \dau collisions, branching ratio of the parent particle, 
\g-conversion factor etc.

\begin{table}[t]
\caption[]{
Relative systematic uncertainties in percent for the \kshort meson 
measurement. The given ranges indicate the variation of the systematic 
uncertainty over the \pt range of the measurement.
}
\label{table_syserr_kshort} 
\begin{ruledtabular}\begin{tabular}{lccc} 
Source                       & \dau      &    \cucu    & Uncertainty  \\
                             & (\%)      & (\%)        &   Type      \\ 
\hline       
Raw yield                    &  4--31     &  14--26   & A       \\
extraction                   &            &           &    \\ 
Acceptance                   &  6          &  5         & B       \\
ERT                          &  2--7      &  3--4     & B       \\
efficiency                   &            &           &   \\
EMCal energy                 &  4--5      &  3--6     & B       \\
resolution                   &            &           & \\
EMCal scale                  &  4--5      &  3--5     & B       \\
\piz selection               &  5--11     &  6--10    & B       \\
$\gamma$ conversion          &  6          &  6         & C       \\
Branching ratio              &  0.2        &  0.2       & C       \\
BBC cross section            &  8        & --        & C        \\
\end{tabular}\end{ruledtabular} 
\end{table} 

The main systematic uncertainty of the \kstar measurement include 
uncertainties in the raw yield extraction, EMCal-PC3 matching, TOF PID 
cuts, track momentum reconstruction, acceptance and BBC cross section. The 
systematic uncertainty associated with the raw yield extraction is 
estimated by varying the fitting ranges, varying the width of the \kstar 
meson peak by $\pm$2\% around its simulated value and taking the integral 
of the fitted RBW function instead of summing up the yield in each \pt 
bin. In addition, the yield difference when the \kstar meson mass is fixed 
to the PDG value and when it is a free parameter in the fit of the mass 
spectrum, is included in the systematic uncertainty. To evaluate the 
uncertainties from EMCal-PC3 matching and TOF PID cuts, the corresponding 
cuts are varied within $\pm$17\%.  The uncertainty in momentum 
reconstruction is estimated by varying the momentum scale within 0.5\% in 
the simulation. A summary of the systematic uncertainties for the case of 
``kaon identified'' analysis technique in \pp, \dau and \cucu collisions 
is given in Table~\ref{table_syserr_pp}.

\begin{table}[b]
\caption[]{Relative systematic uncertainties in percent for the \kstar 
meson measurement in ``kaon identified'' technique. 
The given ranges indicate the variation of the systematic uncertainty over the \pt range of the measurement.} 
\label{table_syserr_pp} 
\begin{ruledtabular}\begin{tabular}{lcccc} 
Source                  & \pp      & \dau    &    \cucu   & Uncertainty  \\
                        &          & (MB)    & (MB)       & Type \\
                        &   (\%)   & (\%)    & (\%)       & \\
 \hline       
Raw yield               & 5--8    & 7--12  &  2--4   & A      \\
extraction              &         &        &         &   \\
Acceptance              & 1--5    & 3--7   &  1--3   & B     \\
Track Momentum          & 1--4    & 2--7   &  1--5   & B      \\
reconstruction          &         &        &          & \\
Track Matching                & 1--4    & 4--7   &  2--13  & B       \\
TOF PID                     & 1--6    & 4--9   &  1--4   & B     \\
BBC cross section       & 10      & 8     &   --    & C \\
\end{tabular}\end{ruledtabular} 
\end{table}

\section{Results and Discussions}
\label{sec:Results}

In this section we present \pt spectra of \kshort and \kstar mesons in 
\pp, \dau and \cucu collisions at \sqsn = 200 GeV. The invariant \pt 
spectra are used to calculate the nuclear modification factors in \dau and 
\cucu collisions at different centralities. These nuclear modification 
factors are compared to those previously measured for neutral pions, 
charged kaons, $\phi$ mesons and protons.

\subsection{Invariant transverse momentum spectra}
\label{pTspectra}

\begin{figure}[thb]
\includegraphics[width=1.0\linewidth]{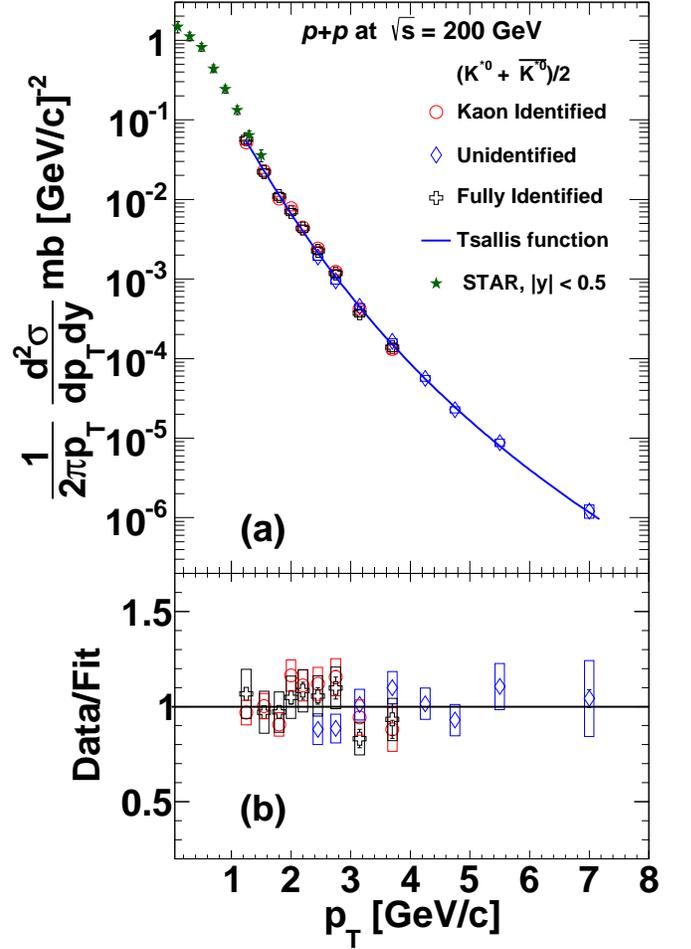} 
\caption{(color online)
(a) \kstar meson invariant yield as a function of \pt obtained with the 
``kaon identified'', ``fully identified'' and ``unidentified'' analysis 
techniques in \pp collisions at \energypp. The systematic uncertainties 
shown with boxes are mostly uncorrelated between analysis techniques. The 
solid blue line is the Tsallis function fit to the combined data points. 
The star symbols are the \kstar meson measurements from the STAR 
collaboration~\cite{starKstarpp}. (b) Ratio of the yields obtained with 
the three analysis techniques to the fit function. The scale uncertainty 
of 10\% is not shown.
}
\label{fig:pppt}
\end{figure}

\begin{figure*}[htb]
\includegraphics[width=0.998\linewidth]{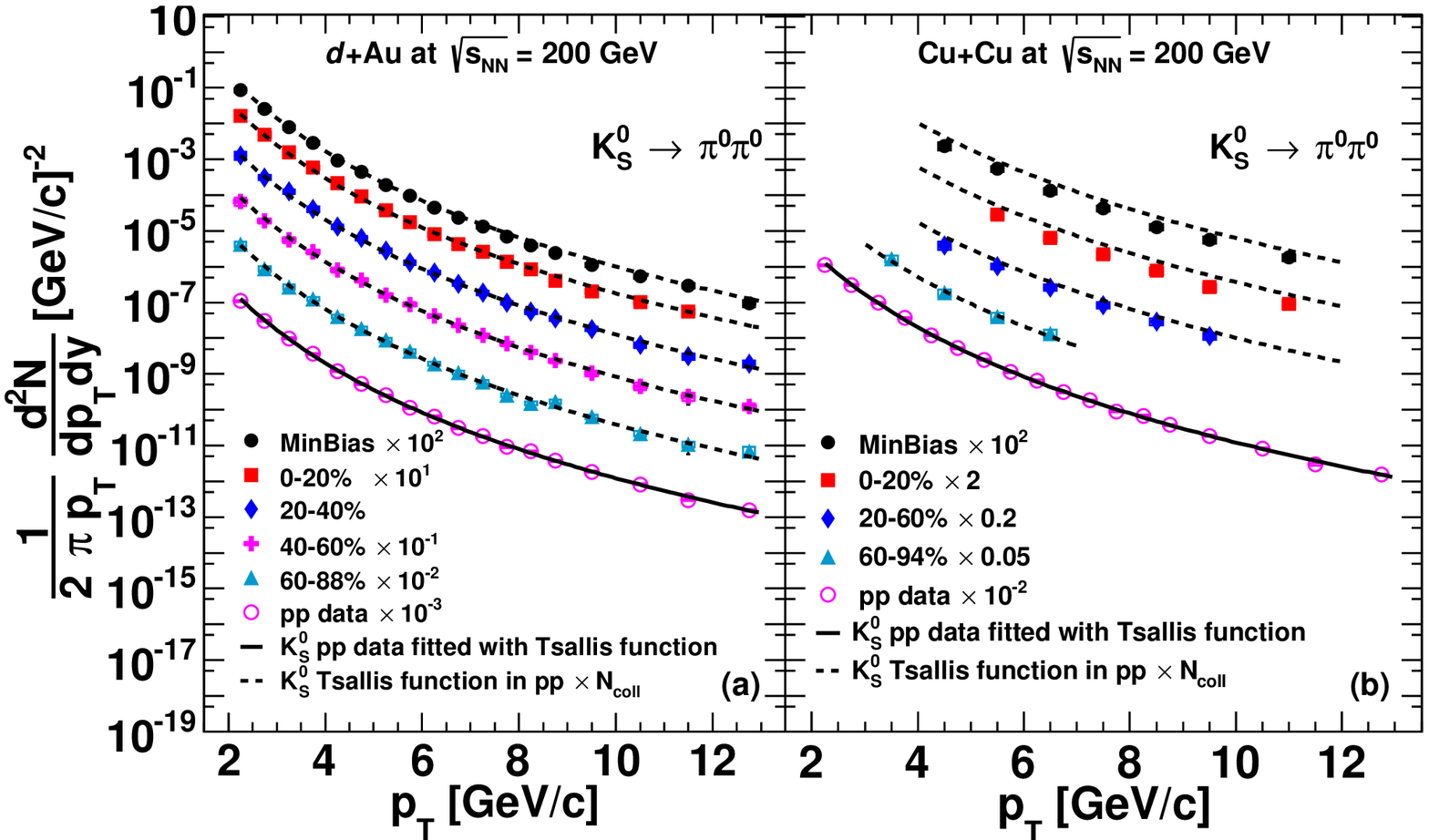} 
\caption{(color online)
\kshort meson invariant \pt spectra (a) for \dau and (b) for \cucu 
collisions at \energy for different centrality bins. The systematic 
uncertainties are shown by the boxes. The solid curves are a fit of the 
\kshort \pp data by the Tsallis function~\cite{PPG099}. The dashed curves 
are the fit function scaled by \Nc. The global \pp uncertainty of $\sim$ 
10\% is not shown.
}
\label{fig:kscucupt}

  \includegraphics[width=0.998\linewidth]{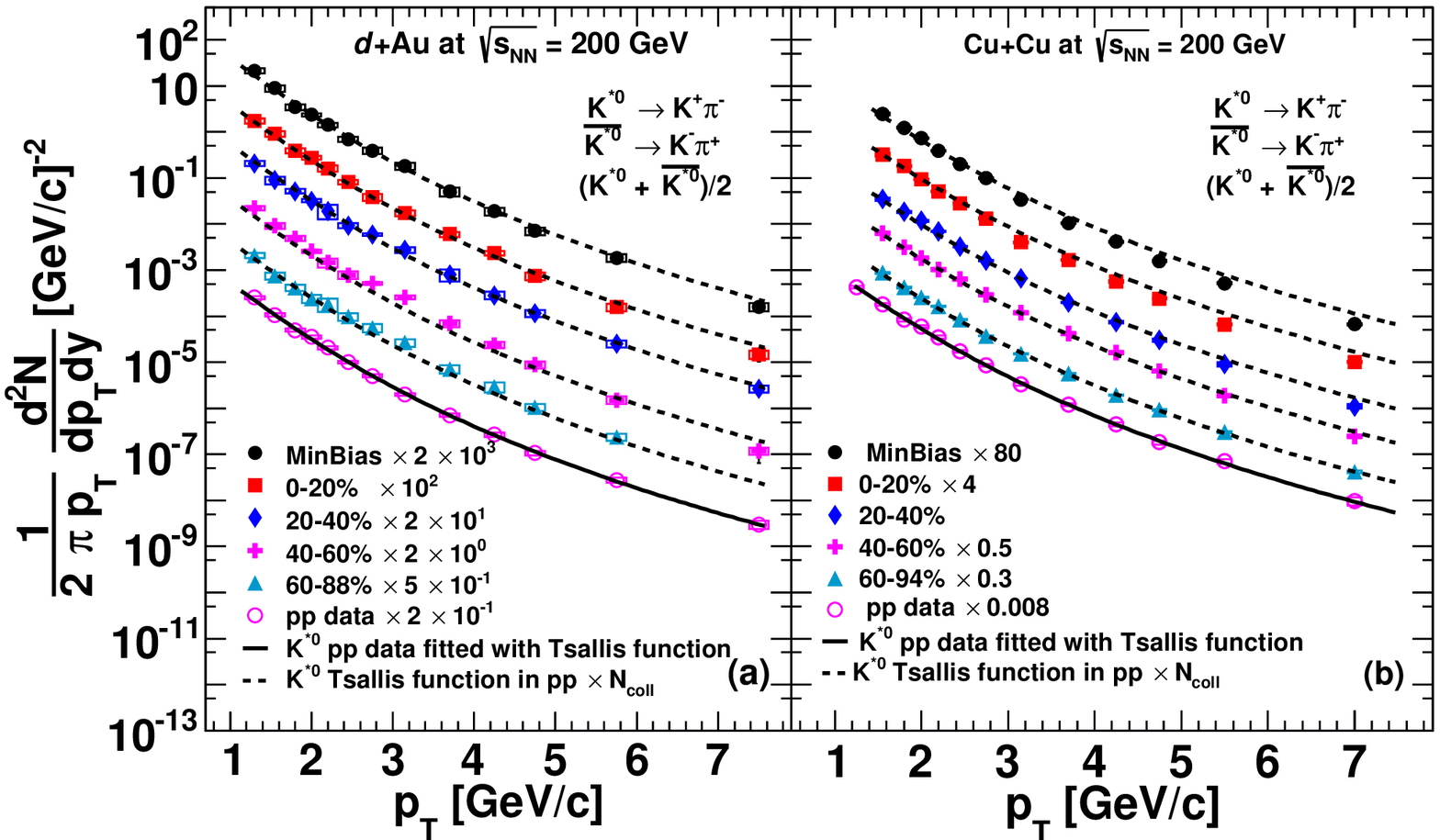} 
  \caption{(color online)
\kstar meson invariant \pt spectra (a) for \dau and (b) for \cucu 
collisions at \energy for different centrality bins. The systematic 
uncertainties are shown by the boxes. The solid curve is a fit of the 
\kstar \pp data by the Tsallis function~\cite{PPG099}. The dashed curves 
are the fit function scaled by \Nc. The global \pp uncertainty of $\sim$ 
10\% is not shown.
}
  \label{fig:all_cent_cucu}
\end{figure*}

Figure~\ref{fig:pppt}~(a), shows the invariant yield of \kstar mesons as a 
function of \pt in \pp collisions at \sqs = 200 GeV. Experimental points 
shown with different symbols correspond to the different analysis 
techniques listed in Table~\ref{sampletype}. The systematic uncertainties, 
mostly uncorrelated for different techniques, are shown along with the 
data points and include raw yield extraction, track matching and TOF PID 
uncertainties listed in Table~\ref{table_syserr_pp}. 


\begin{figure*}[tbh]
  \includegraphics[width=0.47\linewidth]{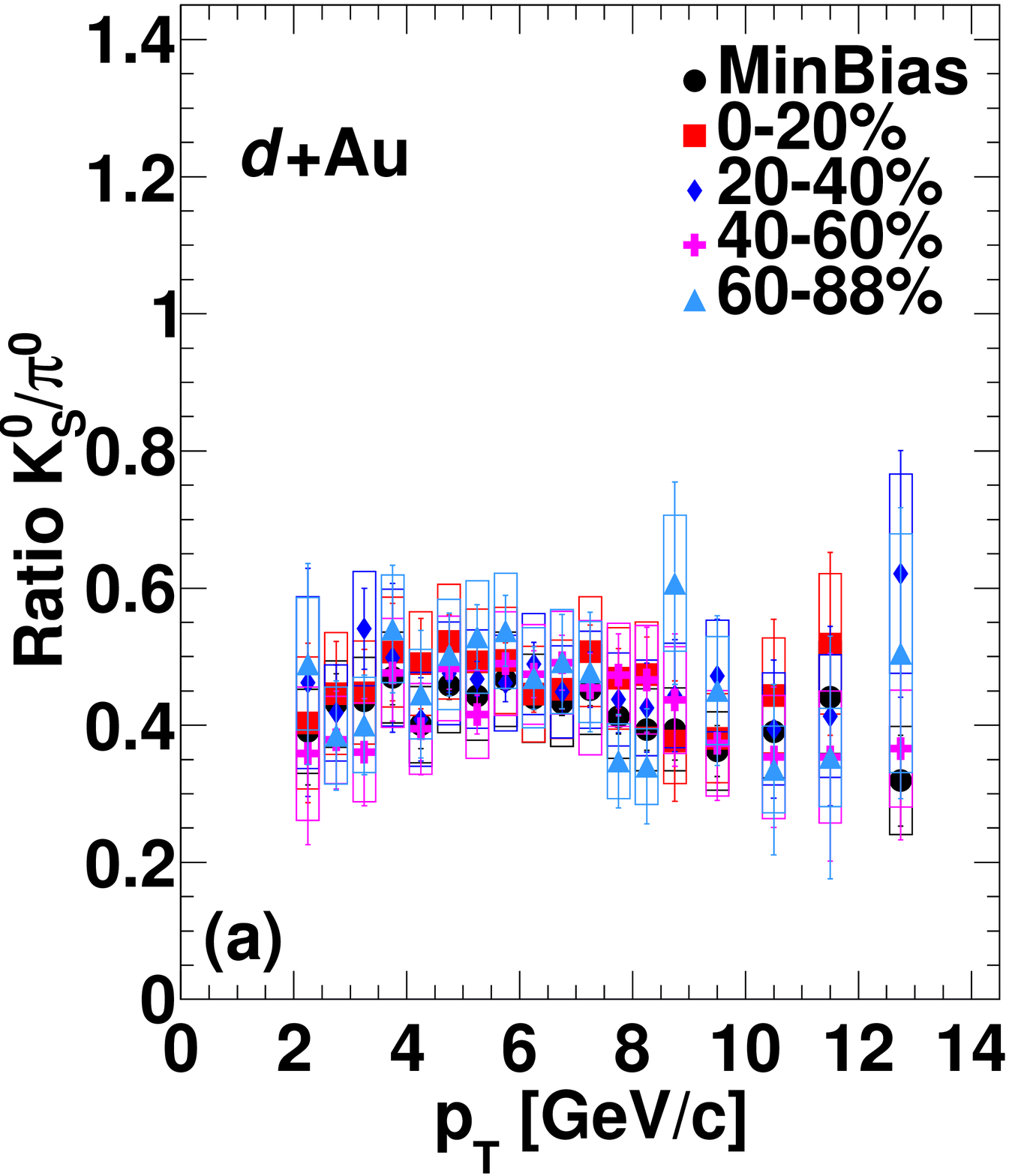}
  \includegraphics[width=0.47\linewidth]{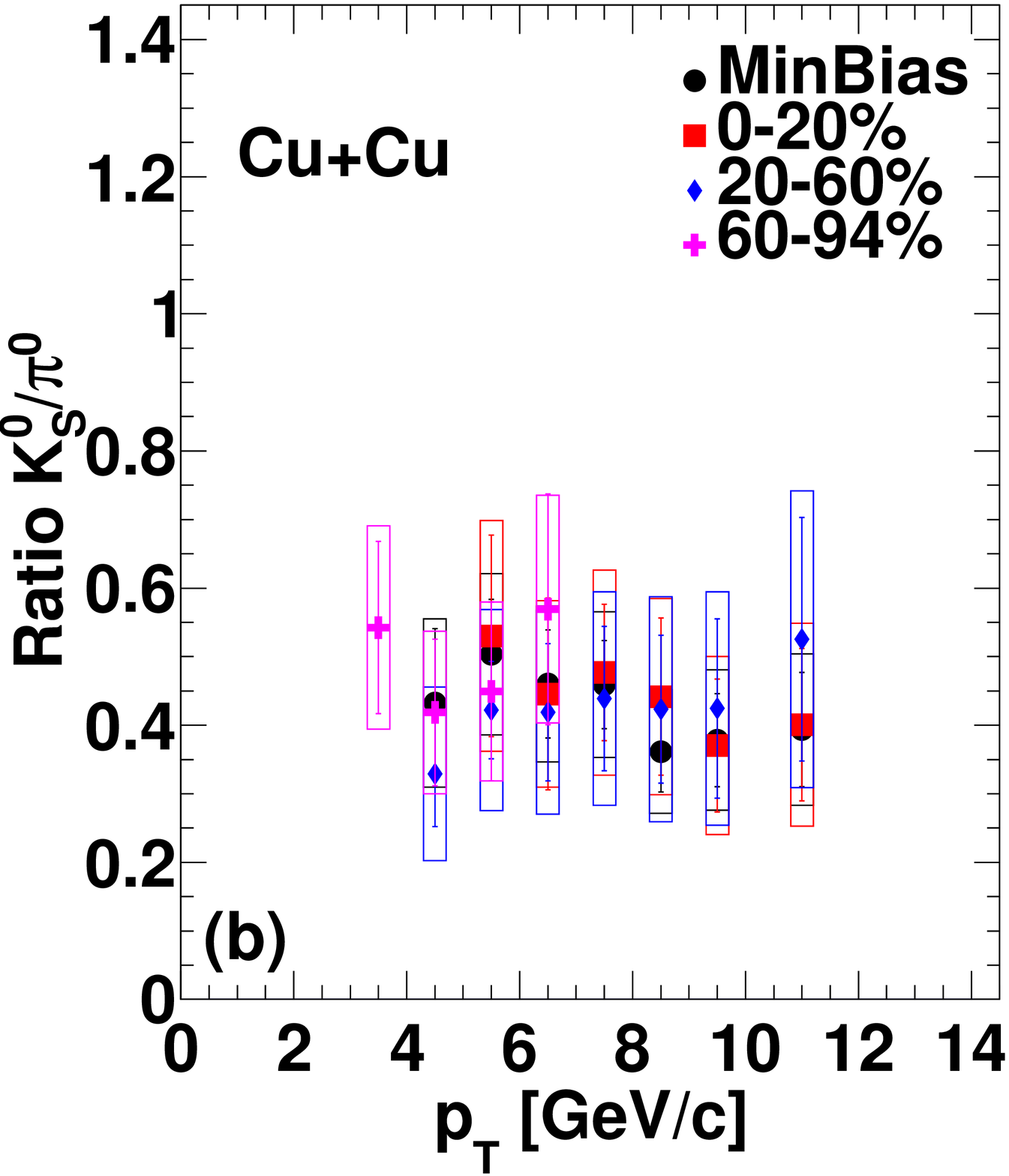}
  \caption{(color online)
\kshort/\piz ratios for (a) \dau and (b) \cucu collisions at \energy for 
different centrality bins. The statistical uncertainties are shown by 
vertical bars and the systematic uncertainties are shown by the boxes.}
  \label{fig:ks_pi0_ratio}
\end{figure*}

\begin{table}[hb]
\caption{\Nc and \Np in \dau and \cucu collisions at \energy.} 
\label{ncoll_npart} 
\begin{ruledtabular}\begin{tabular}{cccccc} 
& Collisions &   Centrality bin (\%) & $\langle\Nc\rangle$ & $\langle\Np\rangle$ \\
\hline
& \dau  & 0--20  &  15.1 $\pm$ 1.0 & 15.3 $\pm$ 0.8 & \\
&       & 20--40 &  10.2 $\pm$ 0.7 & 11.1 $\pm$ 0.6 & \\
&       & 40--60 &  6.6 $\pm$ 0.4  & 7.8  $\pm$ 0.4 & \\
&       & 60--88 &  3.1 $\pm$ 0.2  & 4.3  $\pm$ 0.2 & \\
&       & 0--100 &  7.6 $\pm$ 0.4  & 8.5  $\pm$ 0.4 & \\
\\
& \cucu & 0--20            &  151.8 $\pm$ 17.1 & 85.9 $\pm$ 2.3 & \\
&       & 20--40           &  61.6  $\pm$ 6.6  & 45.2 $\pm$ 1.7 & \\
&       & 40\--60           &  22.3 $\pm$ 2.9 & 21.2 $\pm$ 1.4  & \\
&       & 60--94           &  5.1 $\pm$ 0.7 & 6.4 $\pm$ 0.4     & \\
&       & 0--94            &  51.8 $\pm$ 5.6 & 34.6 $\pm$ 1.2   & \\
&       & 20--60           &  42.0 $\pm$ 4.8 & 33.2 $\pm$ 1.6   & \\
\end{tabular}\end{ruledtabular} 
\end{table} 

The solid line in Figure~\ref{fig:pppt}~(a) is the result of a common fit 
of the data with the Tsallis function in the form used in~\cite{PPG099}:
\begin{eqnarray}
\label{formula3}
\frac{1}{2\pi}\frac{d^2N}{dydp_T} & = & \frac{1}{2\pi} \frac{dN}{dy} \frac{(n-1)(n-2)}{(nT+m(n-1))(nT+m)}   \nonumber \\
                                  & \times & \left(\frac{nT+m_T}{nT+m}\right)^{-n},
\end{eqnarray} 
where $dN/dy$, $n$, and $T$ are the free parameters, 
$m_T=\sqrt{\pt^2 + m^2}$ and $m$ is the mass of the particle of interest. 
The parameter $T$ determines the shape of the spectrum at low \pt where 
particle production is dominated by soft processes whereas $n$ governs the 
high \pt part of the spectrum dominated by particles produced in hard 
scattering. The fit parameters to the \pp data are $dN/dy$ = 1.28 $\pm$ 
0.14, $T$ = 121.077 $\pm$ 19.17 (MeV) and $n$ = 9.67 $\pm$ 0.62 with 
$\chi^{2}/NDF$ = 6.9/10. The uncertainties in the parameters include both 
the statistical and systematic uncertainties in quadrature. 
Figure~\ref{fig:pppt}~(b) shows the ratio of the \kstar meson yields 
obtained with the different techniques to the fit. A good agreement is 
observed for the yields obtained with different analysis techniques that 
demonstrates the robustness of the results. The final \kstar production 
spectrum is obtained by standard weighted averaging~\cite{PDG} of the 
yields and uncorrelated errors for the same \pt bin obtained from the 
different analysis techniques.  The STAR experiment measured 
the \kstar over the \pt range 0--1.5\,\gevc, shown by the solid star 
symbols in Fig.~\ref{fig:pppt}~(a). In the overlap region STAR results 
agree with our measurement within one sigma of combined statistical and 
systematic uncertainties.

Figures~\ref{fig:kscucupt} and \ref{fig:all_cent_cucu} show the invariant 
\pt spectra of \kshort and \kstar mesons in \dau and \cucu collisions at 
\energy. The results for different centrality bins are scaled by arbitrary 
factors for clarity. The final \pt spectrum for \kstar meson in \pp 
collisions at $\sqrt{s}=200$~GeV is shown by the magenta-colored open 
circles in Fig.~\ref{fig:all_cent_cucu}~(b). The \pp results for \kshort, 
both the data points and the Tsallis fit, are taken from 
Ref.~\cite{PPG099}. The solid curves represent the Tsallis fit to the \pp 
data. The dashed curves represent the same fit, scaled by the number of 
binary collisions corresponding to the centrality bins concerned. In \dau 
collisions, the production of both mesons follows the binary scaling for 
all centralities in the measured \pt range. A similar behavior is also 
observed in peripheral \cucu collisions. In central and semi-central \cucu 
interactions, the production of \kshort and \kstar mesons is suppressed at 
$p_T>4$\,\gevc and $p_T>2$--3\,\gevc, respectively.

\begin{figure*}[thb]
  \begin{minipage}{0.5\linewidth}
\includegraphics[width=0.998\linewidth]{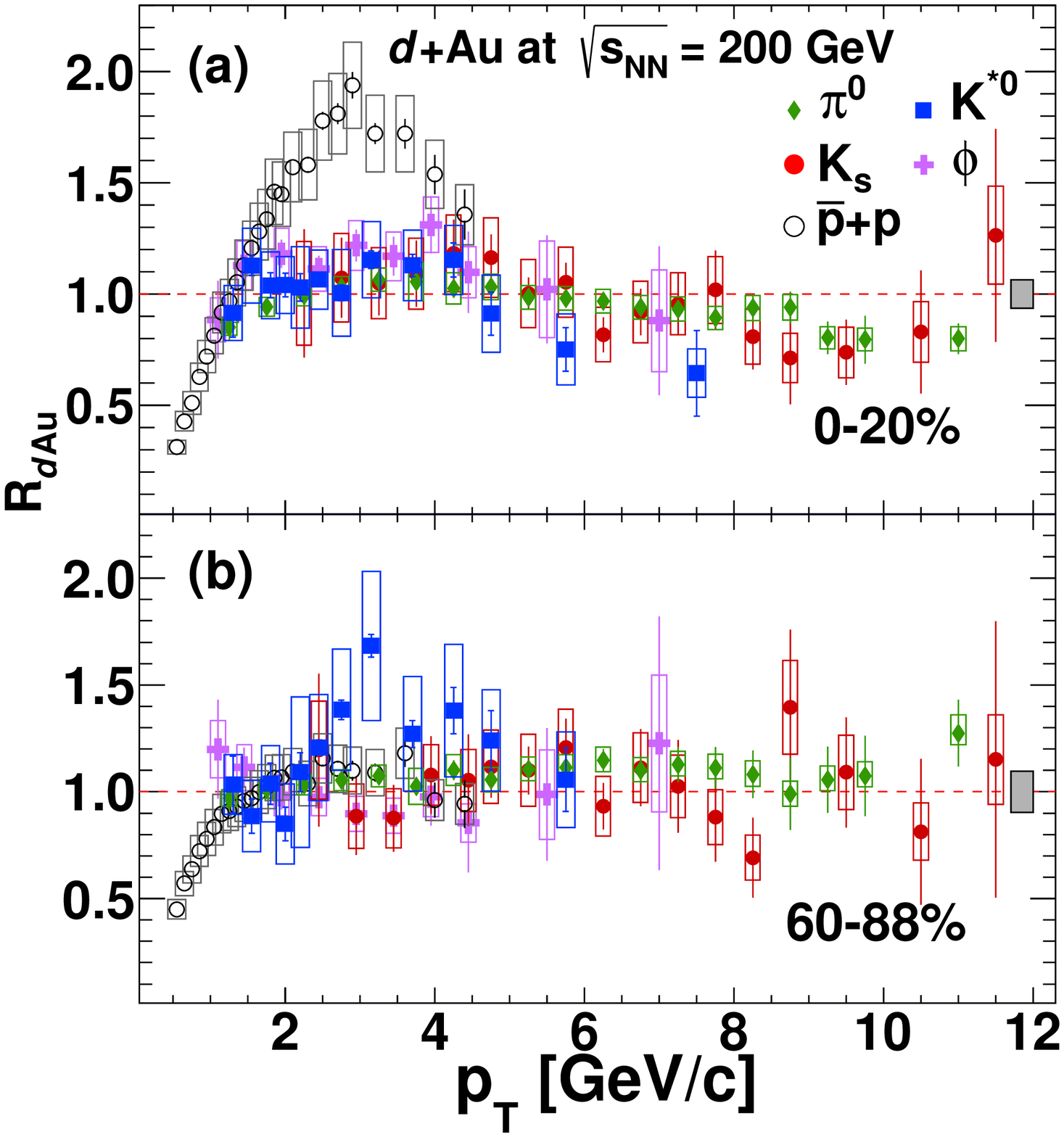} 
  \end{minipage}
  \hspace{0.05\linewidth}
  \begin{minipage}{0.3\linewidth}
\caption{\label{fig:dau_selected} (color online)
Nuclear modification factor as a function of \pt for \kshort and \kstar 
for (a) most central and (b) most peripheral \dau collisions at \energy. 
Results from \piz~\cite{triggerbiaspp}, $\phi$~\cite{phipaper} and 
protons~\cite{PPG146} are also shown.  The \piz results are shown from the 
data collected in 2003 and the results of the rest of the particles are 
obtained from 2008 data. The corresponding systematic uncertainties are 
shown by boxes. The global \pp uncertainty of $\sim$ 10\% is not shown.
}
  \end{minipage}
\includegraphics[width=0.998\linewidth]{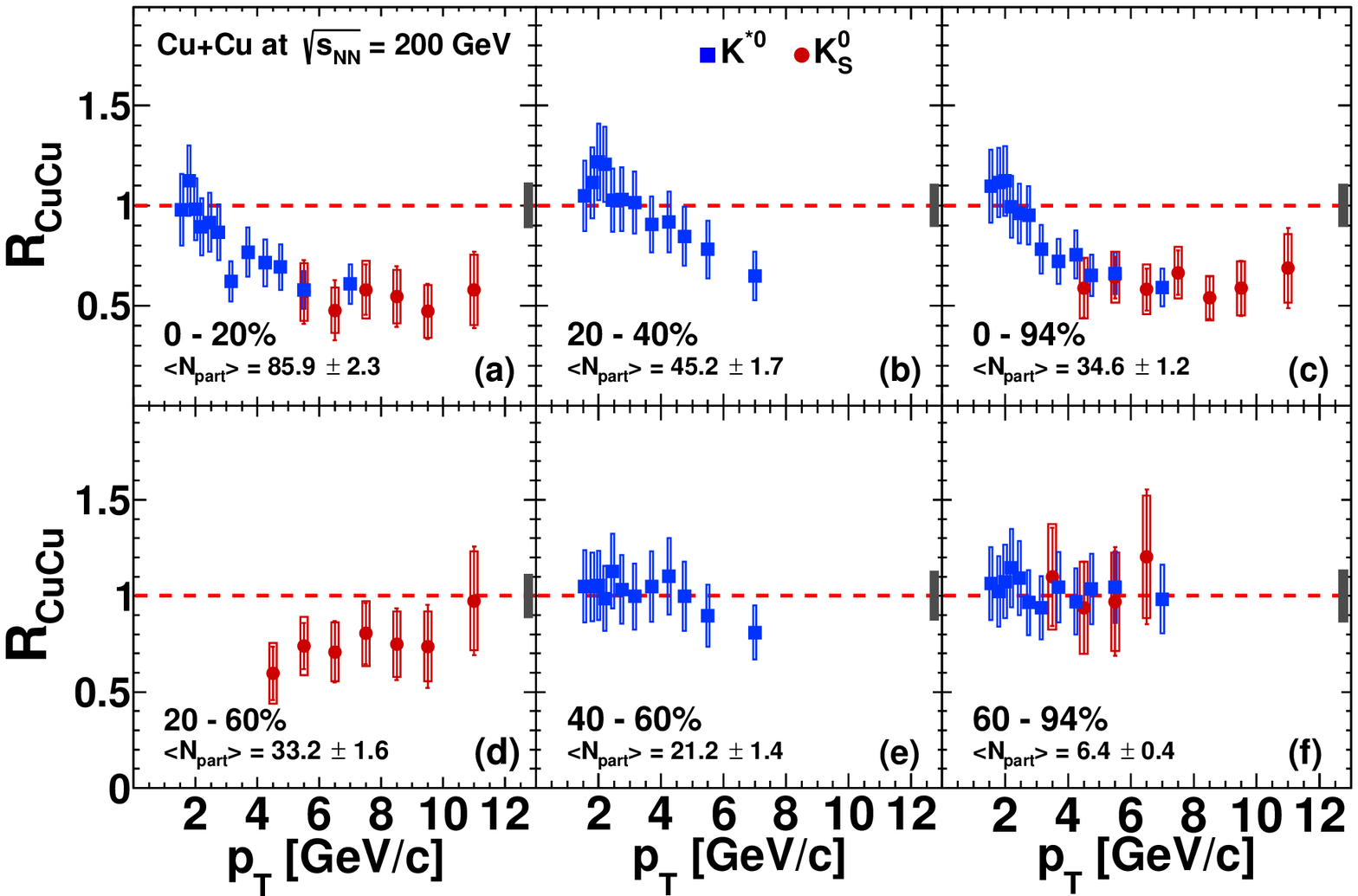} 
\caption{\label{fig:R_CuCu_phi_kstar} (color online)
The nuclear modification factor as a function of \pt for \kshort and 
\kstar meson for centrality bins (a) 0\%--20\% ($\langle\Np\rangle$ = 85.9 
$\pm$ 2.3),
(b) 20\%--40\% ($\langle\Np\rangle$ = 45.2 $\pm$ 1.7), 
(c) 0\%--94\% ($\langle\Np\rangle$ = 34.6 $\pm$ 1.2),
(d) 20\%--60\% ($\langle\Np\rangle$ = 33.2 $\pm$ 1.6), 
(e) 40\%--60\% ($\langle\Np\rangle$ = 21.2 $\pm$ 1.4) and 
(f) 60\%--94\% ($\langle\Np\rangle$ = 6.4 $\pm$ 0.4) in \cucu collisions
at \sqsn = 200 GeV. 
In all panels the statistical uncertainties are shown with vertical bars 
and the systematic uncertainties are shown with boxes. The global \pp 
uncertainty of $\sim$ 10\% is not shown.
}
\end{figure*}



Figure~\ref{fig:ks_pi0_ratio} shows the ratio $\kshort/\piz$ for different 
centrality bins in \cucu collisions at \energy. The ratio 
is flat with respect to \pt with a value of $\sim$ 0.5, irrespective of 
the system and collision centrality. The statistical uncertainties are 
shown by vertical bars and the systematic uncertainties are shown by 
boxes.


\subsection{Nuclear Modification Factors}
\label{R_aa}

The nuclear modification factors for \kshort and \kstar mesons were 
calculated using Eq.~\ref{eq:rAA}. The average number of inelastic 
nucleon-nucleon collisions $\langle\Nc\rangle$ and participants 
$\langle\Np\rangle$ estimated for each centrality bin analyzed in \dau and 
\cucu collisions are summarized in Table~\ref{ncoll_npart}~\cite{glauber, 
klaus_glauber}.

Figure~\ref{fig:dau_selected} shows the nuclear modification factors \rda, 
measured for the \kshort and \kstar mesons in the most central and 
peripheral \dau collisions at \energy. Within uncertainties, the \rda are 
consistent with unity for all centralities at $p_T>1$\,\gevc. However, in 
the most central \dau collisions, there is a hint of a modest Cronin-like 
enhancement in the range $2<p_T<5$\,\gevc and of suppression at 
$p_T>6$--8\,\gevc. Results for $\phi$ and \piz mesons~\cite{triggerbiaspp, 
phipaper} and protons~\cite{PPG146} are also shown for comparison in 
Fig.~\ref{fig:dau_selected}. The \rda for all measured mesons shows 
similar behavior. Based on these results one can conclude that either the 
CNM effects do not play an important role in the production of these 
mesons or different CNM effects compensate each other in the studied \pt 
range. Unlike mesons, baryons~\cite{PPG146} exhibit a strong enhancement 
at intermediate transverse momenta in (semi)central \dau collisions that 
could be explained by recombination models~\cite{reco}.


\begin{figure*}
\begin{minipage}{0.49\linewidth}
\includegraphics[width=0.998\linewidth]{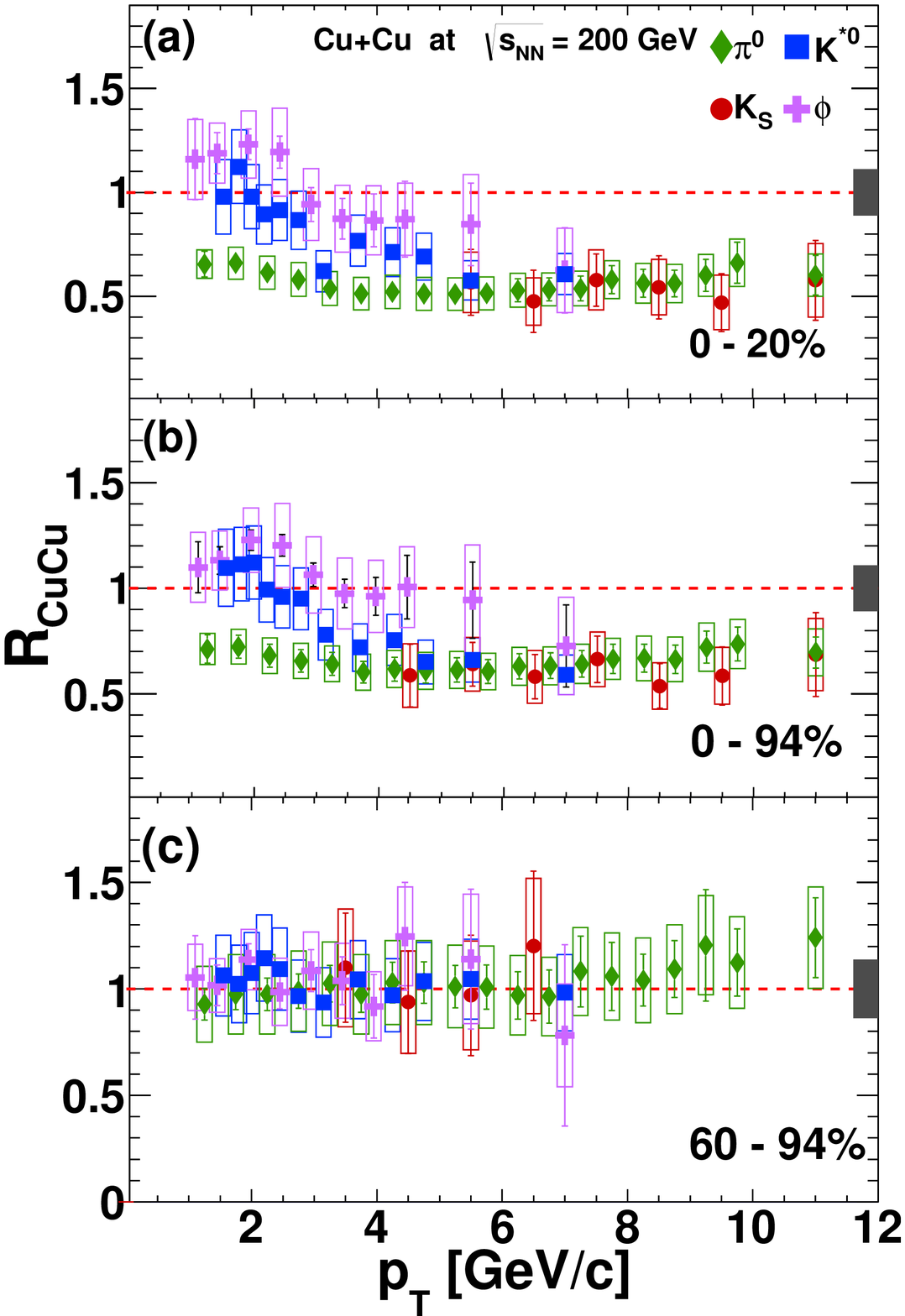}
\caption{(color online)
Nuclear modification factor as a function of \pt for \kshort, \kstar for 
centralities (a) 0\%--20\%, 
(b) 0\%--94\% (MB) and 
(c) 60\%--94\% in  \cucu collisions at \energy.  
Results from \piz~\cite{PPG084pi0cucu} and $\phi$~\cite{phipaper} are also 
shown. The statistical errors are shown by vertical bars. The systematic 
uncertainties are shown by boxes. The global \pp uncertainty of $\sim$ 
10\% is not shown.
}
\label{fig:r_cucu}
\end{minipage}
\begin{minipage}{0.49\linewidth} 
\includegraphics[width=0.998\linewidth]{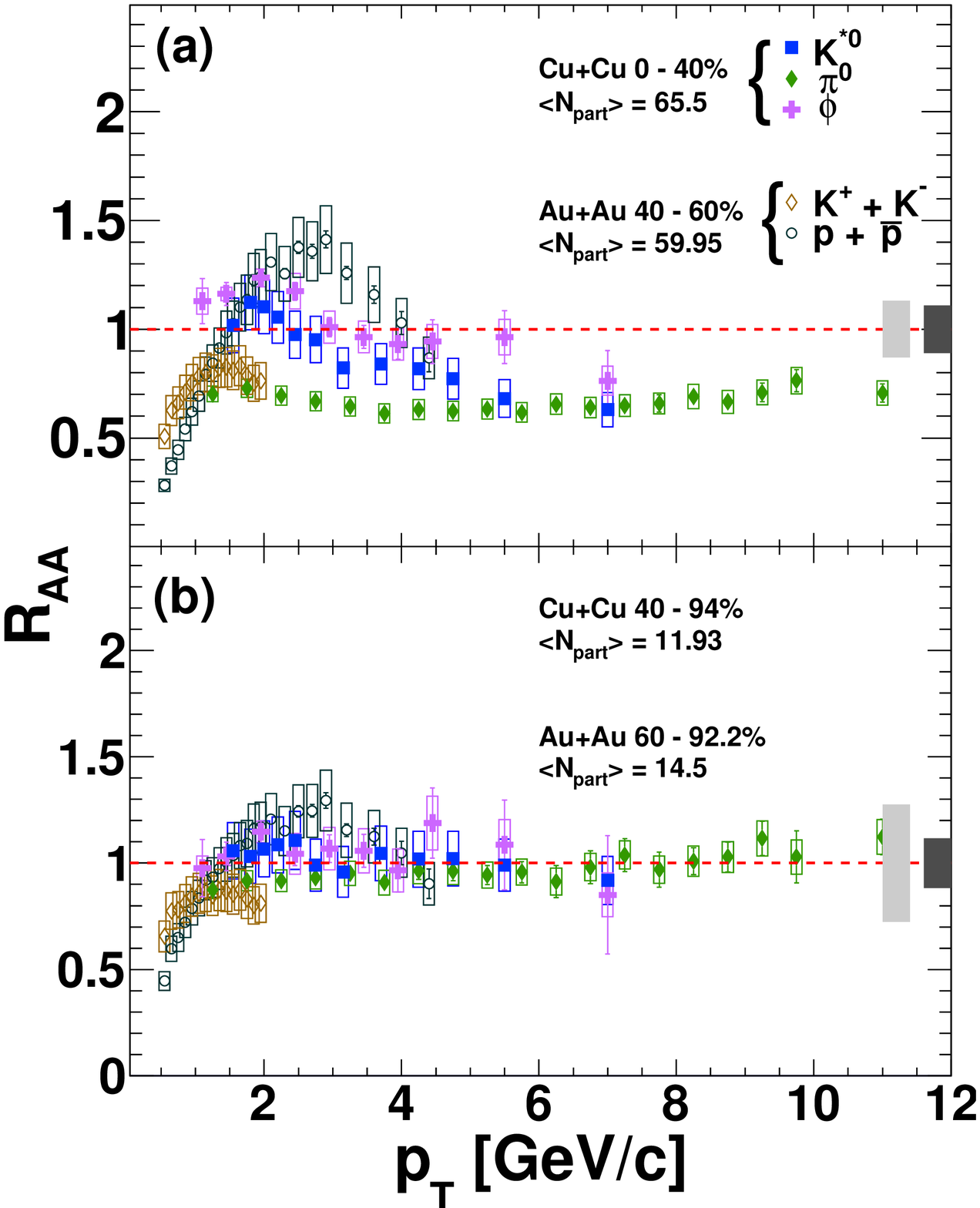}
\caption{(color online)
Comparison of the nuclear modification factor of 
\piz~\cite{PPG084pi0cucu}, $\phi$~\cite{phipaper}, and \kstar in \cucu 
collisions and proton~\cite{PPG146} and kaon~\cite{PPG146} in \auau 
collisions at \energy. The comparisons are made for (a) 40\%--60\% and (b) 
60\%--92\% in \auau system and 0\%--40\% and 40\%--94\% in the \cucu system 
corresponding to similar \Np values in the two systems. The statistical 
errors are shown by vertical bars.  The systematic uncertainties are shown 
by boxes. The global \pp uncertainty of $\sim$ 10\% is not shown.
}
\label{fig:r_cucu_auau}
\end{minipage}
\end{figure*}

Figure~\ref{fig:R_CuCu_phi_kstar} shows the nuclear modification factors 
\rcucu measured for \kshort and \kstar meson in \cucu collisions at 
\energy.  The results are presented for different centrality bins 
corresponding to the $\langle\Nc\rangle$ and $\langle\Np\rangle$ given in 
Table~\ref{ncoll_npart}.  In peripheral \cucu collisions the production of 
\kshort and \kstar mesons follows the binary scaling as expected from 
Figs.~\ref{fig:kscucupt} and~\ref{fig:all_cent_cucu}. The \rcucu factors 
become smaller with increasing centrality and in the most central \cucu 
collisions the production of both mesons is suppressed.  For the most 
central collisions, \rcucu reaches a value of 0.5 at $p_T>5$\,\gevc, both 
for \kshort and \kstar mesons, for the most central collisions.

Figure~\ref{fig:r_cucu} compares the \rcucu results for \kshort and \kstar 
mesons to results obtained for the \piz meson~\cite{PPG084pi0cucu} and 
$\phi$ meson~\cite{phipaper} in the most central, most peripheral, and MB 
\cucu collisions. In peripheral collisions, the nuclear modification 
factors are consistent with unity for all measured mesons at all \pt. In 
central and MB collisions, above \pt $\ge$ 5\,\gevc, the \rcucu of all 
mesons is below unity, and within the uncertainties the suppression is the 
same for all measured mesons, indicating that its mechanism does not 
depend on the particle species. However, at lower \pt between 1--5\,\gevc, 
there are differences among the different particles. The \kstar meson 
\rcucu shows no suppression at \pt $\sim$ 1--2\,\gevc and then decreases 
with increasing \pt, as previously observed for the $\phi$ meson. The \piz 
meson shows significantly stronger suppression over the same \pt range.

Figure~\ref{fig:r_cucu_auau} compares the suppression patterns of 
light-quark mesons, strange mesons, and baryons.  Shown are the \raa of 
\piz, \kstar and $\phi$ mesons measured in \cucu at \energy.  Because 
there are no measurements of \raa for protons and charged kaons in the \cucu 
system, we compare to proton and charged kaon measurements made in \auau 
collisions at the same energy~\cite{PPG146}. The comparisons are made for 
centrality bins corresponding to similar number of participating nucleons 
(\Np), in the \cucu and \auau systems: \cucu 40\%--94\% 
($\langle\Np\rangle$ = 11.93 $\pm$ 0.63) and \auau 60\%--92\% 
($\langle\Np\rangle$ = 14.5 $\pm$ 2.5) in the bottom panel and \cucu 
0\%--40\% ($\langle\Np\rangle$ = 65.5 $\pm$ 2.0) and \auau 40\%--60\% 
($\langle\Np\rangle$ = 59.95 $\pm$ 3.5) in the top panel. In peripheral 
collisions the \raa factors for all mesons are consistent with unity at 
$p_T>2$\,\gevc. A modest enhancement of $\approx$\,1.3 is observed for 
protons. In central collisions, all hadrons show suppression. In the 
intermediate \pt range (\pt = 2--5 \gevc), there seems to be some 
hierarchy with baryons being enhanced, neutral pions being suppressed the 
most and \kstar and $\phi$ mesons showing an intermediate behavior. At 
higher \pt, all particles are suppressed and they seem to reach the same 
level of suppression, within uncertainties, irrespective of their mass or 
quark content. The fact that \raa of all mesons becomes the same is 
consistent with the assumption that energy loss occurs at the parton level 
and the scattered partons fragment in the vacuum. We also note that the 
\raa of the \kstar and $\phi$ mesons appear to be very similar to the \raa 
of electrons from the semi-leptonic decay of heavy flavor mesons
~\cite{SINGE2}. The present results provide additional constraints to the 
models attempting to quantitatively reproduce the nuclear modification 
factors in terms of energy loss of partons inside the medium.

\section{Summary and Conclusions}
\label{sec:Summary}

The PHENIX experiment measured \kshort and \kstar meson production via 
$\piz\piz$ and $K^{\pm}$$\pi^{\mp}$ decay, respectively, in \pp, \dau and 
\cucu collisions at \energy. The invariant transverse momentum spectra and 
nuclear modification factors are presented for different centralities in 
the \dau, and \cucu systems covering the \pt range of 1.1--8.5\,\gevc and 
3--13\,\gevc for \kstar and \kshort respectively. In the \dau system, the 
nuclear modification factor of \kshort and \kstar mesons is almost 
constant as a function of \pt and consistent with unity showing that cold 
nuclear matter effects do not play a significant role in the measured 
kinematic range. A similar behavior is seen in $R_{d\rm{Au}}$ for all 
measured mesons. In the \cucu collisions system, no nuclear modification is 
registered in peripheral collisions within the uncertainties of the 
measurement. In central \cucu collisions both mesons show suppression. In 
the range \pt = 2-5\,\gevc, the strange mesons show an intermediate 
suppression between the more suppressed \piz and the nonsuppressed 
baryons. This behavior provides a particle species dependence of the 
suppression mechanism and provides additional constraints to the models 
attempting to quantitatively reproduce nuclear modification factors. At 
higher \pt, all particles, \piz, strange mesons and baryons, show a 
similar level of suppression.

  
\section*{ACKNOWLEDGMENTS}


We thank the staff of the Collider-Accelerator and Physics
Departments at Brookhaven National Laboratory and the staff of
the other PHENIX participating institutions for their vital
contributions.  We acknowledge support from the 
Office of Nuclear Physics in the
Office of Science of the Department of Energy,
the National Science Foundation, 
Abilene Christian University Research Council, 
Research Foundation of SUNY, and
Dean of the College of Arts and Sciences, Vanderbilt University 
(U.S.A),
Ministry of Education, Culture, Sports, Science, and Technology
and the Japan Society for the Promotion of Science (Japan),
Conselho Nacional de Desenvolvimento Cient\'{\i}fico e
Tecnol{\'o}gico and Funda\c c{\~a}o de Amparo {\`a} Pesquisa do
Estado de S{\~a}o Paulo (Brazil),
Natural Science Foundation of China (P.~R.~China),
Ministry of Science, Education, and Sports (Croatia),
Ministry of Education, Youth and Sports (Czech Republic),
Centre National de la Recherche Scientifique, Commissariat
{\`a} l'{\'E}nergie Atomique, and Institut National de Physique
Nucl{\'e}aire et de Physique des Particules (France),
Bundesministerium f\"ur Bildung und Forschung, Deutscher
Akademischer Austausch Dienst, and Alexander von Humboldt Stiftung (Germany),
OTKA NK 101 428 grant and the Ch. Simonyi Fund (Hungary),
Department of Atomic Energy and Department of Science and Technology (India), 
Israel Science Foundation (Israel), 
National Research Foundation of Korea of the Ministry of Science,
ICT, and Future Planning (Korea),
Physics Department, Lahore University of Management Sciences (Pakistan),
Ministry of Education and Science, Russian Academy of Sciences,
Federal Agency of Atomic Energy (Russia),
VR and Wallenberg Foundation (Sweden), 
the U.S. Civilian Research and Development Foundation for the
Independent States of the Former Soviet Union, 
the Hungarian American Enterprise Scholarship Fund,
and the US-Israel Binational Science Foundation.



\end{document}